# Thermo-viscous feedback in heat transport through fracture–matrix systems with finite thermal capacity

Alessandro Lenci *

*Department of Civil, Chemical, Environmental, and Materials Engineering (DICAM), University of Bologna, Viale del Risorgimento 2, 40136, Bologna, Italy*
*Department of Energy Science & Engineering, Stanford University, 94305, Stanford, CA, USA*
*Univ Rennes, CNRS, Géosciences Rennes, UMR 6118, 35042, Rennes Cedex, France*

ABSTRACT

In geothermal systems, injected fluid flows along preferential fracture pathways and exchanges heat with matrix blocks. Reduced models often neglect temperature-dependent viscosity and assume unlimited matrix storage, although the thermal capacity accessible to the flow is finite. This work analyzes thermo-viscous feedback in a planar fracture–matrix system where viscosity controls discharge through the integrated hydraulic resistance, while finite capacity enters through an effective single-timescale memory fixed by initial uptake and total capacity. The reduced model gives parsimonious semi-analytical predictions with physically based parameters, verified against the nonlinear solution. A perturbation hierarchy separates matrix storage from the hydraulic response it induces, while an exponential velocity reconstruction preserves the exact terminal discharge and bounds front-moment errors within the front-moment limit model. At fixed pressure difference, heating accelerates and broadens the front, whereas cooling delays and sharpens it. The terminal discharge depends only on the endpoint viscosity ratio, while matrix uptake, storage and contrast determine its approach time. At fixed rate, viscosity alters only the pressure drop, not the thermal response. Finite thickness controls the transition to capacity-limited equilibration, and strong matrix uptake collapses the isoviscous response onto a two-variable similarity surface.

## 1. Introduction

Fluid circulation through fractured rocks controls heat transport in enhanced geothermal systems (EGS), subsurface thermal-energy storage (Fleuchaus et al., 2018), groundwater remediation, and thermal-tracer experiments. Fractures commonly provide the dominant flow paths, whereas the surrounding matrix stores most of the thermal energy. Heat is therefore advected rapidly along fractures and exchanged conductively with a slower matrix, so predictions of thermal breakthrough must represent both processes (Viswanathan et al., 2022). Field experiments confirm that matrix conduction attenuates and delays thermal signals relative to solute transport (Hawkins et al., 2017; de La Bernardie et al., 2018; Hoffmann et al., 2022).

The classical analytical description represents a fracture as a lower-dimensional conduit coupled to transverse matrix diffusion (Lauwerier, 1955; Bodvarsson, 1969; Gringarten et al., 1975). Eliminating the matrix field makes the exchange nonlocal in time: the interfacial flux depends on the history of the wall temperature through a geometry-dependent memory kernel (Haggerty and Gorelick, 1995; Carrera et al., 1998; Babey et al., 2015). Semi-infinite and bounded kernels are established for solute and heat transport (Carslaw and Jaeger, 1959; Tang et al., 1981; Sudicky and Frind, 1982; Roubinet et al., 2012; Zhou et al., 2017, 2019), including fracture arrays and variable block sizes (Bödvarsson and Tsang, 1982; Abbasi et al., 2017; Zhou and Wang, 2023). This framework captures the early diffusion-controlled response, but the semi-infinite approximation fails once the thermal penetration depth becomes comparable with the available rock thickness.

In a parallel-fracture array, the symmetry plane between adjacent fractures acts as a no-flux boundary and defines a matrix slab whose thickness is half the intervening rock thickness (Gringarten et al., 1975; Sudicky and Frind, 1982; Bödvarsson and Tsang, 1982). As diffusion spans the slab, the matrix progressively equilibrates with the injected fluid, limiting further conductive exchange. Finite block size consequently governs the late-time tail (Zhou et al., 2017; Volponi et al., 2025), while longitudinal thermal dispersion can modify both transient and steady responses (Zhou et al., 2022; Dejam and Hassanzadeh,

* Correspondence to: Department of Civil, Chemical, Environmental, and Materials Engineering (DICAM), University of Bologna, Viale del Risorgimento 2, 40136, Bologna, Italy.
*E-mail address:* alessandro.lenci@unibo.it.

2023). Roughness and aperture variability introduce additional channeling, residence-time dispersion, and non-Fickian behavior (Neuville et al., 2010; Fox et al., 2015; Klepikova et al., 2021; De Simone et al., 2023; Lenci et al., 2026b); these effects lie outside the uniform-fracture baseline considered here.

A separate limitation of most reduced descriptions is that the fracture velocity is prescribed independently of temperature. Water viscosity can change substantially over the temperature contrasts of geothermal reinjection models and heat-tracer tests (Fox et al., 2015; Guo et al., 2016; Hawkins et al., 2017; de La Bernardie et al., 2018; Hoffmann et al., 2022). At fixed flow rate, this change is carried by the required pressure drop. Under fixed pressure difference, temperature, global hydraulic resistance and heat transport form a feedback loop. In heated or cooled channels (Ockendon and Ockendon, 1977; Richardson, 1987), variable viscosity produces multiple pressure–flow branches and fingering when the walls act as a permanent heat source or sink (Wylie and Lister, 1995; Helfrich, 1995). Detailed geothermal simulations include variable viscosity alongside spatial flow redistribution, deformation or buoyancy (Bataillé et al., 2006; Okoroafor and Horne, 2022; Liu et al., 2024), which makes the isolated hydraulic feedback difficult to identify. Shear-rate-dependent rheology likewise couples viscosity to channeling and fracture–matrix heat transport (Lenci et al., 2022, 2024; Lenci and Daprà, 2026); here viscosity depends on temperature and acts through the global hydraulic resistance.

A gap therefore remains between analytical theories that resolve conductive memory but usually assume constant fluid properties and these coupled numerical models. Even when an analytical treatment solves for the flow field rather than prescribing it Fox et al. (2016), temperature-independent properties close the hydraulic problem without feedback from the evolving thermal state. To isolate this coupling, we combine finite conductive memory with temperature-dependent hydraulic resistance in a reduced theory for a planar fracture embedded in a finite matrix, considering heating and cooling under fixed pressure difference and fixed flow rate. The analysis asks how viscosity contrast shifts breakthrough, how finite matrix thickness controls front advance and spreading, and which hydraulic observable carries the coupling under each operating protocol. A tempered effective memory represents finite diffusive exchange through the initial uptake and total capacity, yielding elementary crossover laws. The exact planar-slab memory, at comparable numerical cost, provides the geometry-specific comparison. An implicit nonlinear solution is complemented by a perturbation expansion in the logarithmic viscosity contrast, whose first-order problem is driven by a scalar resistance functional. An exponential reconstruction of the velocity history controls front-moment errors through the velocity error in a front-moment limit model with rapid transverse equilibration, negligible longitudinal conduction and an untruncated downstream domain (Sections 3.4 and 4.3.2; Appendix C).

Four results follow. First, the integrated hydraulic resistance links viscosity to discharge, with a terminal velocity fixed by the injected-to-initial viscosity ratio under fixed pressure difference. Second, matrix diffusion and capacity, together with the contrast itself, set the timescales over which that endpoint is approached. Third, the operating protocol selects the hydraulic observable affected by viscosity: discharge at fixed pressure difference, the fracture pressure drop at fixed rate. Fourth, under strong matrix uptake the isoviscous response collapses onto a two-variable similarity surface, on which the tempered and exact closures share the late advance rate but differ in breakthrough timing, tails and front spreading. The framework is a controlled baseline for separating these mechanisms from geometrical, mechanical, buoyant, and multiphase effects and for subsequent upscaling (Lenci et al., 2026a).

Section 2 defines the fracture–matrix problem, governing equations, hydraulic closure and dimensionless controls. Section 3 presents the shared matrix-memory closures, the nonlinear reference solution, the perturbation hierarchy and the exponential velocity reconstruction that preserves the terminal discharge. Section 4 specifies the reference configuration, defines the transport observables and assesses sensitivity to the viscosity law, contrast truncation and memory closure. Section 5 quantifies hydraulic feedback, breakthrough, energy partition and front spreading, examines finite-matrix similarity, and compares fixed-pressure and fixed-rate operation. Section 6 discusses the physical implications, assumptions and practical applications, and Section 7 summarizes the conclusions.

Appendix A derives the semi-infinite closed forms and their thermoviscous correction. Appendix B derives the finite-matrix similarity and crossover laws and quantifies differences between the memory closures. Appendix C establishes the front-moment representation and the bounds linking velocity errors to moment errors. Appendix D documents the numerical implementation, verification tests, computational benchmark and literature inputs.

## 2. Physical problem and governing equations

### 2.1. Fracture–matrix heat transport

Consider a planar fracture of length $L$ and half-aperture $b$, bounded by two identical rock-matrix domains of thickness $h_m$. By symmetry, the problem is formulated over the half-domain $0 < z < b$ in the fluid and $b < z < b + h_m$ in the matrix (Fig. 1). Here $x$, $z$, and $t$ denote the longitudinal coordinate, transverse coordinate, and time, respectively. Let $u(t)$ denote the section-mean fracture velocity and $u_0$ its value in the initial constant-viscosity state.

The temperature fields are normalized as

$$\Theta_i(x, z, t) = \frac{T_i(x, z, t) - T_0}{T_{\text{inj}} - T_0}, \qquad i \in \{f, m\}, \tag{2.1}$$

where $T_f$ and $T_m$ are the fluid and matrix temperatures, $T_0$ is the initial temperature, and $T_{\text{inj}}$ is the imposed inlet temperature; write $\Delta T = T_{\text{inj}} - T_0$.

Let $\kappa_{Lf}$ and $\kappa_{Tf}$ denote the longitudinal and transverse fluid thermal diffusivities. Heat transport in the fracture follows the macroscopic two-dimensional advection–diffusion model

$$\frac{\partial \Theta_f}{\partial t} + u(t)\frac{\partial \Theta_f}{\partial x} = \kappa_{Lf}\frac{\partial^2 \Theta_f}{\partial x^2} + \kappa_{Tf}\frac{\partial^2 \Theta_f}{\partial z^2}, \qquad 0 < x < L, \quad 0 < z < b. \tag{2.2}$$

The local velocity profile determines hydraulic resistance through its Poiseuille weighting, whereas heat is advected with the section-mean velocity, as in macroscopic fracture–matrix formulations (Tang et al., 1981; Sudicky and Frind, 1982; Roubinet et al., 2012). The hydraulic feedback is introduced in Section 2.2. The closure rests on fast transverse diffusion, measured by the Graetz number $Gz = (b^2/\kappa_{Tf})/(L/u_0)$; Section 4.1 reports its range in the computed cases and the error of the closure estimated with the resolved velocity profile. Longitudinal spreading induced by the transverse velocity profile, including its coupling to matrix heat exchange, requires separate upscaling and is not resolved here (Dejam and Hassanzadeh, 2023). Molecular longitudinal conduction is retained through $\kappa_{Lf}$.

Let $\kappa_{Lm}$ and $\kappa_{Tm}$ denote the longitudinal and transverse matrix thermal diffusivities. The matrix temperature satisfies

$$\frac{\partial \Theta_m}{\partial t} = \kappa_{Lm}\frac{\partial^2 \Theta_m}{\partial x^2} + \kappa_{Tm}\frac{\partial^2 \Theta_m}{\partial z^2}, \qquad 0 < x < L, \quad b < z < b + h_m. \tag{2.3}$$

The matrix is impermeable, so heat is transported there by conduction alone. The evaluated configurations are thermally isotropic, $\kappa_{Lf} = \kappa_{Tf} = \kappa_f$ and $\kappa_{Lm} = \kappa_{Tm} = \kappa_m$; the directional notation is retained in the governing equations to show where anisotropy would enter.

For density $\rho_i$ and specific heat $c_{p,i}$ ($i = f, m$), $\Gamma_i = \rho_i c_{p,i}$ is the volumetric heat capacity of the fluid or saturated matrix, with capacity ratio $\sigma = \Gamma_m/\Gamma_f$. Temperature and heat flux are continuous at the fracture–matrix interface, $\Theta_f(x, b, t) = \Theta_m(x, b, t)$, with

$$\kappa_{Tf}\left.\frac{\partial \Theta_f}{\partial z}\right|_{z=b} = \sigma\kappa_{Tm}\left.\frac{\partial \Theta_m}{\partial z}\right|_{z=b}. \tag{2.4}$$

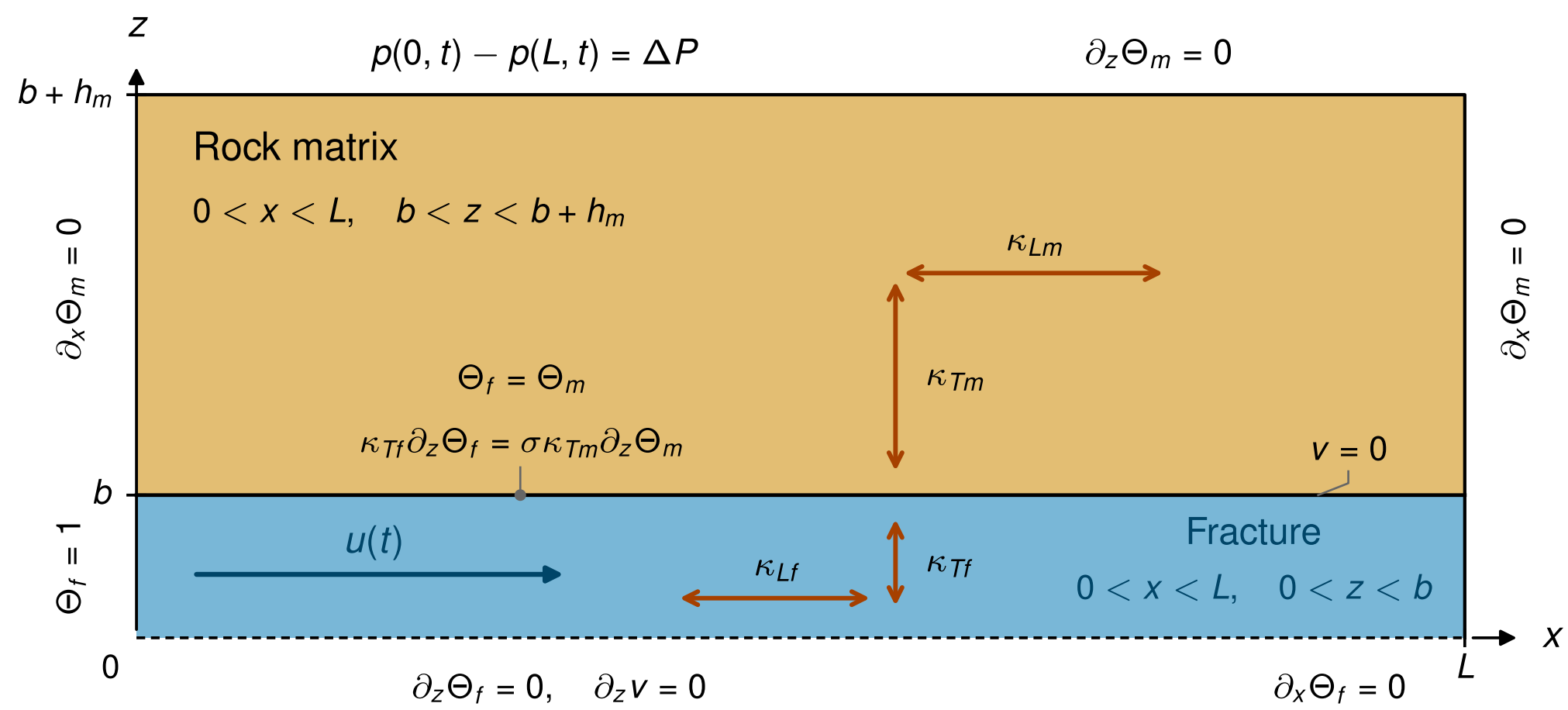

**Fig. 1.** Half-domain of the fracture–matrix system, with geometry and hydraulic and thermal boundary conditions (not to scale). Blue: advection at the velocity $u(t) = u_0 U(t)$ set by the temperature field through Eq. (2.19) at fixed pressure difference; orange: thermal diffusion.

Eq. (2.4) expresses continuity of the conductive flux, with conductivities $k_i = \Gamma_i \kappa_{Ti}$. Initially, $\Theta_f(x,z,0) = \Theta_m(x,z,0) = 0$. The fracture inlet is subjected to a unit temperature step, $\Theta_f(0,z,t) = 1$ for $t > 0$, while the fracture outlet satisfies

$$\left.\frac{\partial \Theta_f}{\partial x}\right|_{x=L} = 0. \tag{2.5}$$

Symmetry at the fracture mid-plane requires

$$\left.\frac{\partial \Theta_f}{\partial z}\right|_{z=0} = 0. \tag{2.6}$$

The matrix satisfies

$$\left.\frac{\partial \Theta_m}{\partial x}\right|_{x=0} = \left.\frac{\partial \Theta_m}{\partial x}\right|_{x=L} = 0, \tag{2.7}$$

and an insulated outer boundary,

$$\left.\frac{\partial \Theta_m}{\partial z}\right|_{z=b+h_m} = 0. \tag{2.8}$$

### 2.2. Temperature-dependent hydraulic response

The fluid is incompressible and Newtonian, with dynamic viscosity $\mu(T_f)$. Density and thermal properties are constant and gravity is neglected, so temperature affects the flow only through viscosity. The fracture is driven by a fixed inlet-to-outlet pressure difference $\Delta P$; the fixed-rate protocol is considered separately in Section 5.8. We assume laminar flow and use a quasi-steady hydraulic closure, with momentum and pressure adjustment fast relative to thermal evolution. The viscous adjustment time $b^2/\nu_{\text{kin},0}$, with $\nu_{\text{kin},0} = \mu(T_0)/\rho_f$ the reference kinematic viscosity, is about 0.25 s in the reference configuration. The discharge is determined by the current temperature field.

Let $v(x,z,t)$ denote the local streamwise fluid velocity. Under the lubrication approximation, the streamwise momentum balance is

$$\frac{\partial}{\partial z}\left[\mu(T_f)\frac{\partial v}{\partial z}\right] = \frac{\partial p}{\partial x}, \tag{2.9}$$

where $p = p(x,t)$ is the pressure to leading order. Symmetry at the fracture mid-plane and no slip at the wall require

$$\left.\frac{\partial v}{\partial z}\right|_{z=0} = 0, \qquad v(x,b,t) = 0. \tag{2.10}$$

Integrating Eq. (2.9) twice across the half-aperture, with the conditions of Eq. (2.10), yields

$$v(x,z,t) = -\frac{\partial p}{\partial x}\int_z^b \frac{z'}{\mu[T_f(x,z',t)]}\,\mathrm{d}z'. \tag{2.11}$$

Because of symmetry, the cross-sectionally averaged velocity introduced in Eq. (2.2) can be evaluated over the half-aperture,

$$u(t) = \frac{1}{b}\int_0^b v(x,z,t)\,\mathrm{d}z = -\frac{1}{b}\frac{\partial p}{\partial x}\int_0^b \frac{z^2}{\mu[T_f(x,z,t)]}\,\mathrm{d}z. \tag{2.12}$$

For later use, define the normalized Poiseuille-weighted hydraulic average of a transverse field $w(z)$ as

$$\langle w\rangle_H = \frac{3}{b^3}\int_0^b z^2 w(z)\,\mathrm{d}z. \tag{2.13}$$

Introducing the reference viscosity $\mu_0 = \mu(T_0)$ and the local mobility ratio

$$m_\mu(\Theta_f) = \frac{\mu_0}{\mu(T_f)}, \tag{2.14}$$

where $T_f$ and $\Theta_f$ are related by Eq. (2.1), the local dimensionless hydraulic mobility is

$$m_H(x,t) = \left\langle m_\mu(\Theta_f)\right\rangle_H = \frac{3}{b^3}\int_0^b z^2 m_\mu\left[\Theta_f(x,z,t)\right]\,\mathrm{d}z. \tag{2.15}$$

For the reference isoviscous state, $\mu = \mu_0$ and therefore $m_H = 1$. Eq. (2.12) can then be written as the local pressure–velocity relation

$$u(t) = -\frac{b^2}{3\mu_0} m_H(x,t)\frac{\partial p}{\partial x}. \tag{2.16}$$

Because the aperture is uniform and the fluid is incompressible, continuity requires the volumetric flux, and hence $u(t)$, to be independent of $x$. The pressure gradient therefore adjusts locally to the temperature-dependent hydraulic resistance, and integration of Eq. (2.16) along the fracture gives

$$\Delta P = \frac{3\mu_0 u(t)}{b^2}\int_0^L \frac{\mathrm{d}x}{m_H(x,t)}. \tag{2.17}$$

The corresponding constant-viscosity reference velocity is

$$u_0 = \frac{b^2 \Delta P}{3\mu_0 L}. \tag{2.18}$$

Write $U(t) = u(t)/u_0$ for the normalized velocity. Under a fixed pressure difference, the hydraulic closure is

$$U(t) = \left[\frac{1}{L}\int_0^L \frac{\mathrm{d}x}{m_H(x,t)}\right]^{-1}. \tag{2.19}$$

Temperature variations across the aperture therefore enter the hydraulic response through the Poiseuille-weighted local mobility $m_H$, whereas longitudinal variations combine through the harmonic average in Eq. (2.19), corresponding to hydraulic resistances in series. The transverse weighting follows from the variable-viscosity momentum balance; it differs from the mobility at the section-mean temperature by a term of order $Gz$, within the accuracy of the section-mean transport closure.

### 2.3. Viscosity model

Define the logarithmic viscosity contrast between the injected and initial states as

$$\lambda = \ln\left[\frac{\mu(T_{\mathrm{inj}})}{\mu(T_0)}\right]. \tag{2.20}$$

Positive $\lambda$ denotes an increase in viscosity and negative $\lambda$ a decrease; for liquid water these correspond to cooling and heating, respectively. The viscosity between these endpoints is approximated by

$$\mu(T_f) = \mu_{\exp}(T_f) = \mu_0 \exp\big(\lambda \Theta_f\big). \tag{2.21}$$

The corresponding mobility ratio is

$$m_\mu(\Theta_f) = \exp\big(-\lambda \Theta_f\big). \tag{2.22}$$

Section 4.4 assesses Eq. (2.21) against the Vogel-type correlation $\mu_V(T) = a_V\, 10^{b_V/(T-c_V)}$, with $T$ in kelvin. For water, $a_V = 2.414 \times 10^{-5}$ Pa s, $b_V = 247.8$ K and $c_V = 140$ K. This correlation reproduces the reference formulation of Huber et al. (2009) within 2% for liquid water near saturation between 20 and 250 °C. The exponential law compresses the temperature dependence into $\lambda$, allowing the perturbation expansion, and matches both endpoint viscosities exactly. Its model-form error is therefore confined to intermediate temperatures.

### 2.4. Finite-matrix response

The finite matrix can be represented through its response to the interfacial heat flux. Introduce the normal coordinate $y = z - b$ and the shifted field $\widetilde{\Theta}_m(x, y, t) = \Theta_m(x, b{+}y, t)$. The normalized signed interfacial heat flux $q_m = Q_m/(\Gamma_f \Delta T)$, where $Q_m$ is the physical flux into the matrix, is

$$q_m(x,t) = -\sigma \kappa_{Tm} \left.\frac{\partial \widetilde{\Theta}_m}{\partial y}\right|_{y=0}, \tag{2.23}$$

and the interfacial temperature is $\Theta_w(x,t) = \widetilde{\Theta}_m(x,0,t) = \Theta_m(x,b,t)$.

Hats denote Laplace-transformed functions. After decomposition into the matrix longitudinal eigenmodes, let $s$ denote the variable conjugate to dimensional time $t$ and $q_x = j\pi/L$, $j = 0, 1, \ldots$, the wavenumbers associated with the no-flux end conditions. The matrix compliance $\widehat{H}_m$ is the wall-temperature response per unit interfacial flux. The transformed relation is

$$\widehat{\Theta}_w(q_x, s) = \widehat{H}_m(q_x, s)\widehat{q}_m(q_x, s), \tag{2.24}$$

where

$$\widehat{H}_m(q_x, s) = \frac{\coth\big(\gamma h_m\big)}{\sigma \kappa_{Tm} \gamma}, \qquad \gamma(q_x, s) = \sqrt{\frac{s + \kappa_{Lm} q_x^2}{\kappa_{Tm}}}. \tag{2.25}$$

The slab response in Eq. (2.25) is the classical bounded-conduction transfer function (Carslaw and Jaeger, 1959); its solute counterpart underlies the analytical fracture–matrix solution of Sudicky and Frind (1982) for a parallel-fracture array, and its semi-infinite limit that of Tang et al. (1981) for a single fracture. Using the partial-fraction expansion of the hyperbolic cotangent,

$$\widehat{H}_m(q_x, s) = \frac{1}{\sigma h_m}\frac{1}{s + \kappa_{Lm} q_x^2} + \frac{2}{\sigma h_m}\sum_{\ell=1}^{\infty} \frac{1}{s + \kappa_{Lm} q_x^2 + \nu_\ell^{\mathrm{ex}}}, \tag{2.26}$$

with the exact-slab (superscript ex) decay rates

$$\nu_\ell^{\mathrm{ex}} = \kappa_{Tm}\left(\frac{\ell \pi}{h_m}\right)^2. \tag{2.27}$$

The zero mode accounts for finite thermal storage, while the decaying modes describe transverse diffusive relaxation.

For longitudinally uniform forcing, $q_x = 0$, the corresponding time-domain kernel is

$$H_m(t) = \frac{1}{\sigma h_m}\left[1 + 2\sum_{\ell=1}^{\infty} \exp(-\nu_\ell^{\mathrm{ex}} t)\right]. \tag{2.28}$$

Its short- and long-time limits are

$$H_m(t) \sim \frac{1}{\sigma\sqrt{\pi \kappa_{Tm} t}}, \qquad t \ll \frac{1}{\nu_1^{\mathrm{ex}}}, \tag{2.29}$$

and

$$H_m(t) \longrightarrow \frac{1}{\sigma h_m}, \qquad t \gg \frac{1}{\nu_1^{\mathrm{ex}}}. \tag{2.30}$$

### 2.5. Characteristic timescales and dimensionless parameters

Table 1 summarizes the physical roles of the principal controls defined below and in Section 2.3.

The reference advective transit time is $t_a = L/u_0$, with dimensionless coordinates $\xi = x/L$ and $\tau = t/t_a$. Histories written as functions of $t$ or $\tau$ refer to corresponding instants; the same convention applies to their perturbation coefficients. The relaxation time of the slowest nonzero transverse ($q_x = 0$) matrix compliance mode is

$$t_1 = \frac{1}{\nu_1^{\mathrm{ex}}} = \frac{h_m^2}{\pi^2 \kappa_{Tm}}. \tag{2.31}$$

The ratio $t_1/t_a$ defines the matrix relaxation number

$$\Pi_1 = \frac{t_1}{t_a} = \frac{u_0 h_m^2}{\pi^2 \kappa_{Tm} L}. \tag{2.32}$$

The wall-exchange kernel that controls breakthrough decays on the timescale $4t_1$ (Appendix B); $t_1$ is retained as the compliance relaxation time. Finite matrix storage is measured by the matrix capacity number

$$C_m = \frac{\sigma h_m}{b}. \tag{2.33}$$

After transverse equilibration, the fracture and matrix capacities act in parallel. For the isoviscous reference state, the equilibrium front speed is $u_0/(1 + C_m)$; this defines the reference capacity-retarded transit time

$$t_r = (1 + C_m) t_a. \tag{2.34}$$

The associated semi-infinite exchange amplitude is

$$\mathcal{A} = \frac{\sigma\sqrt{\kappa_{Tm} t_a}}{b} = \frac{C_m}{\pi\sqrt{\Pi_1}}. \tag{2.35}$$

It measures capacity-weighted matrix penetration over one transit, relative to the half-aperture.

A fracture–matrix exchange Péclet number is used, defined on the full aperture $2b$ and using the effective matrix conductivity $\Gamma_m \kappa_{Tm}$. This retains the independent matrix diffusivity and storage coefficients of the reference parameterization. Because the velocity responds to the temperature field, the number varies with time,

$$Pe(t) = \frac{\Gamma_f\, u(t)\,(2b)}{\Gamma_m \kappa_{Tm}} = \frac{2b\, u(t)}{\sigma \kappa_{Tm}}, \qquad Pe_0 = \frac{2b\, u_0}{\sigma \kappa_{Tm}}. \tag{2.36}$$

Here $Pe_0$ is the initial, isoviscous value. The hydraulic response is therefore reported equivalently as $Pe(t)/Pe_0 = U(t)$. In addition to the Graetz number of Section 2.1, the remaining controls are the half-aperture aspect ratio $\varepsilon$, the matrix diffusivity anisotropy $\alpha_m$, and the fluid-to-matrix transverse diffusivity ratio $\beta_T$:

$$\varepsilon = \frac{b}{L}, \tag{2.37}$$

and

$$\alpha_m = \frac{\kappa_{Lm}}{\kappa_{Tm}}, \qquad \beta_T = \frac{\kappa_{Tf}}{\kappa_{Tm}}. \tag{2.38}$$

**Table 1**
Physical roles of the principal controls. Definitions are given in Section 2.3 and Eqs. (2.32), (2.33) and (2.35); $\mathcal{A}$ is a dependent combination of $C_m$ and $\Pi_1$.

| Control | Physical meaning | Main consequence |
|---|---|---|
| $\lambda$ | Logarithmic viscosity contrast | Sign of hydraulic feedback and terminal discharge at fixed pressure |
| $\mathcal{A}$ | Matrix uptake relative to advection | Early penetration and breakthrough before finite thickness is felt |
| $\Pi_1$ | Matrix relaxation relative to transit | Timing of departure from the semi-infinite response |
| $C_m$ | Matrix-to-fluid thermal capacity | Capacity-retarded front speed after equilibration |

Under thermal isotropy, longitudinal fluid conduction is fixed by the same groups, $u_0 L/\kappa_{Lf} = Gz/\varepsilon^2$, and longitudinal matrix conduction at the front scale is measured by

$$\chi = \frac{\mathcal{A}^2 \varepsilon}{\sigma} = \frac{\sigma \kappa_{Tm}}{u_0 b} = \frac{2}{Pe_0}. \tag{2.39}$$

The exchange Péclet number used in the results is a dependent combination,

$$Pe_0 = \frac{2\pi\sqrt{\Pi_1 Gz\, \beta_T}}{\varepsilon C_m}, \tag{2.40}$$

leaving seven independent controls: $\lambda$, $\Pi_1$, $C_m$, $Gz$, $\varepsilon$, $\alpha_m$ and $\beta_T$. Unless stated otherwise, only $\lambda$, $\Pi_1$ and $C_m$ vary.

## 3. Reference and reduced thermo-viscous formulations

### 3.1. Matrix-memory closures

The memory closure is shared by the nonlinear reference and the reduced formulation, independently of the viscosity-contrast expansion. Superscripts temp and ex label the tempered and exact-slab responses; subscripts $t$, $e$ and $L$ label tempered, exact-slab and semi-infinite Lauwerier kernels, moments and exchange symbols. Tempering supplies an effective memory with elementary crossover laws; the exact slab resolves the planar geometry and supplies the comparisons in Sections 5.6 and 5.7.

The exact finite-slab transfer function is Eq. (2.25) at $q_x = 0$. The tempered form uses the cutoff $t_h = h_m^2/\kappa_{Tm}$:

$$\hat{H}_m^{\mathrm{temp}}(s) = \frac{\sqrt{s + t_h^{-1}}}{\sigma\sqrt{\kappa_{Tm}\, s}}. \tag{3.1}$$

It preserves the early diffusive response, to leading order, and the total matrix capacity: with the dimensionless cutoff $\tau_h = t_h/t_a$, conserving capacity gives $\tau_h = (C_m/\mathcal{A})^2$, which equals $\pi^2 \Pi_1$ for the planar matrix, so tempering adds no free parameter. The strong-uptake crossover is an elementary function of one timescale (Eq. (B.12)). Exponentially truncated memories are used in multirate transfer models (Cvetkovic, 2012; Tecklenburg et al., 2016) and tempered anomalous transport (Meerschaert et al., 2008). The distribution of relaxation times differs from that of the slab and affects variance and breakthrough tails; the spectra and limiting forms are given in Appendix B. At the reference amplitude and $\lambda = 0$, tempering advances half-breakthrough by up to 19% near $\Pi_1 \approx 50$ (Appendix B) and delays the 99% time by up to 32% relative to the exact slab (Table B.1). Timing therefore depends on the closure. A compressed exact-slab representation with a reduced set of relaxation states (Appendix D.4) and tempered memory have comparable computational costs at matched accuracy (Table D.4).

### 3.2. Nonlinear reference solution

The reference solution integrates the governing fracture–matrix equations with the exact hydraulic closure of Eq. (2.19), without an expansion in viscosity contrast. With constant thermal properties, temperature-dependent velocity is the sole nonlinearity. Transverse fracture modes resolve the temperature variation across the aperture, while matrix-memory states retain the effect of past wall heat flux. Their relaxation rates represent diffusion through the slab and replace a resolved matrix grid without discarding its thermal history.

The reference is checked by spatial and temporal refinement, interface continuity, energy conservation and an independent Laplace-domain solution. The discrete system, time integration, settings and verification results are collected in Appendix D.

### 3.3. Perturbation hierarchy

The expansion about the isoviscous state $\lambda = 0$ is

$$\Theta_i = \Theta_i^{(0)} + \lambda \Theta_i^{(1)} + O(\lambda^2), \qquad i \in \{f, m\}, \tag{3.2}$$

$$\frac{u}{u_0} = 1 + \lambda U_1 + O(\lambda^2). \tag{3.3}$$

Superscripts (0) and (1) denote the leading-order field and its first-order coefficient; $U_1(t)$ is the first-order hydraulic response. At this order, the mobility depends only on the leading-order temperature. Define its hydraulic average by

$$A_0(x,t) = \left\langle \Theta_f^{(0)}(x,\cdot,t) \right\rangle_H. \tag{3.4}$$

The leading-order fields satisfy Eqs. (2.2)–(2.4) at $u = u_0$, with the original initial and boundary conditions. They provide both the thermal baseline and the first-order forcing. Expanding Eq. (2.19), with the longitudinal average $\langle f \rangle_x = L^{-1}\int_0^L f(x)\,\mathrm{d}x$, gives

$$U_1(t) = -\langle A_0(\cdot,t)\rangle_x. \tag{3.5}$$

Once the isoviscous field is known, the velocity correction follows from a spatial quadrature. The transverse modal projection used to evaluate $A_0$ is given in Appendix D.1; rapid transverse equilibration makes it close to the section-mean temperature, while the calculations retain the full projection.

The first-order fracture field satisfies

$$\frac{\partial \Theta_f^{(1)}}{\partial t} + u_0 \frac{\partial \Theta_f^{(1)}}{\partial x} = \kappa_{Lf}\frac{\partial^2 \Theta_f^{(1)}}{\partial x^2} + \kappa_{Tf}\frac{\partial^2 \Theta_f^{(1)}}{\partial z^2} - u_0 U_1(t)\frac{\partial \Theta_f^{(0)}}{\partial x}. \tag{3.6}$$

The matrix correction satisfies Eq. (2.3) for $\Theta_m^{(1)}$. Both corrections satisfy the interface conditions and homogeneous initial and boundary data, including $\Theta_f^{(1)}(0,z,t) = 0$. The fracture–matrix operator is unchanged; the sole forcing is the velocity correction in Eq. (3.6).

### 3.4. Finite-contrast reconstruction

The strict first-order thermal approximation is

$$\Theta_i^{(1\mathrm{st})} = \Theta_i^{(0)} + \lambda \Theta_i^{(1)}, \qquad i \in \{f, m\}. \tag{3.7}$$

Its regular remainder is $O(\lambda^2)$, as is that of the linear velocity in Eq. (3.3). When the insulated matrix and fracture equilibrate at the injection temperature, as a semi-infinite matrix also does asymptotically, the resistance integral contains a single viscosity, giving the exact fixed-pressure endpoint

$$U_\infty = \frac{u_\infty}{u_0} = \lim_{t\to\infty}\frac{u(t)}{u_0} = \exp(-\lambda). \tag{3.8}$$

This endpoint is a consistency limit of the hydraulic closure, valid when the fracture carries the resistance between the pressure nodes; it depends only on viscosity contrast, and a permanent thermal sink would change it. The approach depends on contrast and on the competition between diffusive uptake (Eq. (A.5)) and finite-matrix relaxation and storage, measured by $t_1$ and $t_r$ (Appendix D.3).

**Table 2**
Formulations used in the comparisons. Matrix closure is chosen separately (Section 3.1); paired perturbative and nonlinear comparisons retain the same closure and exponential viscosity law. RMS denotes root-mean-square outlet-temperature difference. Quoted ranges quantify reduction error; water-law sensitivity is assessed separately in Section 4.4.

| Formulation | Hydraulic and thermal treatment | Use and tested scope |
|---|---|---|
| Classical reference | Prescribed isoviscous velocity and semi-infinite matrix | Reference for combined viscosity and finite-capacity effects |
| Nonlinear reference | Exact hydraulic closure coupled to the thermal field | Reference at finite contrast; Vogel can be used directly |
| Strict first order | Linear hydraulic and thermal corrections | Outlet RMS $\leq 5\%$, sampled maximum $\leq 23\%$, for $\lvert\lambda\rvert \leq 1$, $\Pi_1 \geq 1$ (Section 5.2); strict moments are nonuniform |
| Exponential reduced model | Second-order velocity reconstruction driving a linear thermal problem | At $\Pi_1 = 10^2$: moment errors below 2% for $\lvert\lambda\rvert \leq 0.6$, up to about 4%/8% (mean/variance) at $\lvert\lambda\rvert = 1$ and 15%/27% at 1.5, both maxima under heating; stronger contrasts require the nonlinear reference |

Let $U_2(t)$ denote the second-order hydraulic coefficient. To reconstruct the endpoint and transient, write

$$\frac{u}{u_0} = 1 + \lambda U_1 + \lambda^2 U_2 + O(\lambda^3). \tag{3.9}$$

Let $A_1 = \langle \Theta_f^{(1)} \rangle_H$ and $B_0 = \langle (\Theta_f^{(0)})^2 \rangle_H$. Expansion of the exact hydraulic closure gives

$$U_2 = \langle A_0 \rangle_x^2 - \langle A_0^2 \rangle_x - \langle A_1 \rangle_x + \tfrac{1}{2} \langle B_0 \rangle_x. \tag{3.10}$$

The longitudinal variance of $A_0$, the first-order field and the cross-aperture second moment determine $U_2$; no second-order temperature solve is required. It vanishes initially and approaches $1/2$, the second-order Taylor coefficient of the terminal velocity.

We define the reconstructed velocity $u_{\text{rec}}$ by

$$\frac{u_{\text{rec}}}{u_0} = \exp\left\{ \lambda U_1 + \lambda^2 \left[ U_2 - \tfrac{1}{2} U_1^2 \right] \right\}. \tag{3.11}$$

It matches Eq. (3.9) through second order, remains positive and reaches the exact endpoint. For fixed physical parameters and matrix closure, the isoviscous and first-order fields are computed once and reused across contrasts. The reconstructed velocity then drives one linear thermal solve per time step, without nonlinear iteration. The field, breakthrough curve and energy partition share this history.

Table 2 distinguishes the formulations and their tested ranges.

## 4. Parameterization and analysis framework

### 4.1. Reference physical configuration

Unless stated otherwise, simulations use the fixed-pressure configuration of Table 3. Internal sweeps vary $\lambda$ and $h_m$; the fixed-rate benchmark of Section 5.8 uses separate inputs.

At each chosen thickness, $\lambda$ varies at fixed remaining dimensionless controls and thermal properties. The dimensional reference uses water properties near 20 °C. For a geothermal temperature pair, retaining $u_0$ requires rescaling $\Delta P$ with the initial-state viscosity according to Eq. (2.18).

Detailed transients use $\lvert\lambda\rvert \leq 0.8$; the error-map grid extends over $-2 \leq \lambda \leq 2.4$. Literature contrasts are listed in Table D.5.

With Eq. (2.35), the Graetz number of Section 2.1 is $Gz = \sigma^2/(\beta_T \mathcal{A}^2)$: for given fluid and rock properties, section-mean advection requires an exchange that is not weak. The reference configuration has $Gz \approx 0.017$, which the thickness sweeps leave unchanged. At fixed pressure the instantaneous value $Gz\, U(t)$ lies between $Gz$ and $Gz\, e^{-\lambda}$: between $1.5 \times 10^{-3}$ and 0.12 over the map grid, between $7.5 \times 10^{-3}$ and 0.037 in the detailed transients, and up to 0.081 in the Vogel tests. Transit times from $t_a/10$ to $10\, t_a$ in the Rotokawa projection (Fig. 5(a)) give $1.7 \times 10^{-3}$–0.17. For the literature cases of Table D.5 whose sources report an aperture and a transit indicator, including the fixed-rate benchmark, order-of-magnitude estimates at the reported conditions give $2 \times 10^{-5} \lesssim Gz \lesssim 0.07$ for planar apertures. An isoviscous profile-resolved check (Appendix D.3, Table D.3) advects heat with the plane-Poiseuille profile. For $Gz \leq 0.3$, the section-mean outlet temperature shifts half-breakthrough by less than 0.6%, while the flux-weighted outlet temperature reaches it earlier, by at most about $17\,Gz$ percent: about 2% at $Gz = 0.12$, the largest $Gz\, U(t)$ of the map grid. Coupled calculations with viscosity-dependent profiles and the exponential law, at the reference $Gz$ and for seven selected contrasts spanning $-2 \leq \lambda \leq 2.4$, change the half-breakthrough of the flux-weighted outlet temperature by at most 0.4%, where it is reached, and that temperature by at most about 1.2% of $\lvert T_{\text{inj}} - T_0 \rvert$.

**Table 3**
Reference fixed-pressure water–rock configuration, derived from Lenci et al. (2026b). In the inherited thermal parameterization, $\kappa_m$ is the solid diffusivity and $\sigma$ uses saturated-matrix capacity; $k_m^{\text{model}} = \Gamma_m \kappa_m$ is the resulting effective model conductivity, not a measured saturated conductivity.

| Symbol | Quantity (unit) | Value |
|---|---|---|
| *Fluid* | | |
| $\rho_f$, $c_{p,f}$ | Density ($\text{kg}\,\text{m}^{-3}$), specific heat ($\text{J}\,\text{kg}^{-1}\,\text{K}^{-1}$) | 1000, 4189 |
| $\Gamma_f$ | Fluid volumetric heat capacity ($\text{J}\,\text{m}^{-3}\,\text{K}^{-1}$) | $4.189 \times 10^6$ |
| $k_f$ | Thermal conductivity ($\text{W}\,\text{m}^{-1}\,\text{K}^{-1}$) | 0.59 |
| $\kappa_f$ | Thermal diffusivity ($\text{m}^2\,\text{s}^{-1}$) | $1.41 \times 10^{-7}$ |
| $\mu_0$ | Reference viscosity (Pa s) | $1.001 \times 10^{-3}$ |
| *Rock matrix* | | |
| $\rho_r$, $c_{p,r}$ | Solid density ($\text{kg}\,\text{m}^{-3}$), specific heat ($\text{J}\,\text{kg}^{-1}\,\text{K}^{-1}$) | 2500, 750 |
| $\Gamma_r$ | Solid volumetric heat capacity, $\rho_r c_{p,r}$ ($\text{J}\,\text{m}^{-3}\,\text{K}^{-1}$) | $1.875 \times 10^6$ |
| $k_r$ | Thermal conductivity of the solid ($\text{W}\,\text{m}^{-1}\,\text{K}^{-1}$) | 3.5 |
| $\kappa_m$ | Matrix thermal diffusivity, $\kappa_m \approx k_r/\Gamma_r$ ($\text{m}^2\,\text{s}^{-1}$) | $1.87 \times 10^{-6}$ |
| $\phi_r$ | Rock porosity | 0.10 |
| $\sigma$ | Saturated-matrix heat-capacity ratio | 0.503 |
| $\Gamma_m$ | Saturated-matrix volumetric heat capacity ($\text{J}\,\text{m}^{-3}\,\text{K}^{-1}$) | $2.11 \times 10^6$ |
| $k_m^{\text{model}}$ | Effective conductivity ($\text{W}\,\text{m}^{-1}\,\text{K}^{-1}$) | 3.94 |
| *Geometry and flow* | | |
| $b$ | Fracture half-aperture (m) | $5.0 \times 10^{-4}$ |
| $L$ | Fracture length (m) | 10 |
| $u_0$ | Reference velocity ($\text{m}\,\text{s}^{-1}$) | $9.406 \times 10^{-2}$ |
| $\Delta P$ | Applied pressure difference (kPa) | 11.3 |
| $t_a$ | Advection time (s) | 106 |
| *Dimensionless groups* | | |
| $Pe_0$ | Fracture–matrix exchange Péclet number | 100 |
| $Gz$ | Graetz number | $1.7 \times 10^{-2}$ |
| $\varepsilon$ | Half-aperture aspect ratio | $5.0 \times 10^{-5}$ |
| $\beta_T$ | Fluid-to-matrix diffusivity ratio | 0.075 |
| $\alpha_m$ | Matrix diffusivity anisotropy | 1 |
| $\mathcal{A}$ | Exchange amplitude, Eq. (2.35) | 14.18 |
| $C_m$ | Matrix capacity number, Eq. (2.33) | Swept |
| $\Pi_1$ | Matrix relaxation number, Eq. (2.32) | Swept |
| $\chi$ | Longitudinal matrix conduction | $2.0 \times 10^{-2}$ |

### 4.2. Finite-matrix thickness and characteristic regimes

The second control variable is the matrix thickness $h_m$. Unlike an analysis in which $\Pi_1$ and $C_m$ are varied independently, changing $h_m$ modifies the matrix relaxation time and thermal storage simultaneously. By Eqs. (2.32) and (2.33), $\Pi_1 \propto h_m^2$ while $C_m \propto h_m$, so that

$$\Pi_1 = \frac{u_0 b^2}{\pi^2 \kappa_{Tm} L \sigma^2} C_m^2 \tag{4.1}$$

for fixed fracture, fluid, and matrix properties.

Variation of $h_m$ therefore traces a physically constrained trajectory in the $(\Pi_1, C_m)$ plane. Small $h_m$ combines rapid transverse equilibration with limited storage, so finite-capacity effects become accessible during breakthrough; at large $h_m$ the thermal disturbance samples only a fraction of the thickness over the advective timescale, approaching the semi-infinite response. At breakthrough, this approximation requires the penetration depth to remain small over the breakthrough time: for the reference amplitude, the exact-slab outlet half-breakthrough time at $\lambda = 0$ is within 1% of its semi-infinite value for $\Pi_1 \gtrsim 10^2$ and about half of it at $\Pi_1 = 10$.

### 4.3. Transport observables

Four observables are monitored: the normalized velocity $U$, the outlet breakthrough curve, the partition of thermal energy, and the first two spatial moments of the fracture temperature. The aperture-averaged fracture temperature used in the latter three is

$$\overline{\Theta}_f(x,t) = \frac{1}{b}\int_0^b \Theta_f(x,z,t)\,\mathrm{d}z. \tag{4.2}$$

It equals the zeroth transverse mode but differs from the Poiseuille-weighted hydraulic average.

The thermal breakthrough curve (BTC) is the aperture-averaged outlet temperature

$$\Theta_{\text{out}}(t) = \overline{\Theta}_f(L,t). \tag{4.3}$$

#### 4.3.1. Thermal-energy partition

The following nonnegative quantities measure the thermal perturbation relative to the initial state: a thermal deficit for cooling and an excess for heating. They are normalized by $\Gamma_f |T_{\text{inj}} - T_0|$ for one fracture half-domain per unit width, so that the contents below have units of area and the boundary fluxes of area per unit time. The fracture and matrix perturbation contents are

$$\mathcal{E}_f(t) = \int_0^L \int_0^b \Theta_f(x,z,t)\,\mathrm{d}z\,\mathrm{d}x, \tag{4.4}$$

$$\mathcal{E}_m(t) = \sigma \int_0^L \int_b^{b+h_m} \Theta_m(x,z,t)\,\mathrm{d}z\,\mathrm{d}x. \tag{4.5}$$

Longitudinal fracture diffusion adds a conductive inlet contribution,

$$Q_{\text{in}}(t) = b\left[u(t) - \kappa_{Lf}\left.\frac{\partial \overline{\Theta}_f}{\partial x}\right|_{x=0}\right], \qquad Q_{\text{out}}(t) = b\,u(t)\Theta_{\text{out}}(t), \tag{4.6}$$

with zero conductive outflow. Their time integrals $\mathcal{E}_{\text{in}}(t) = \int_0^t Q_{\text{in}}(t')\,\mathrm{d}t'$ and $\mathcal{E}_{\text{out}}(t) = \int_0^t Q_{\text{out}}(t')\,\mathrm{d}t'$ satisfy

$$\mathcal{E}_{\text{in}}(t) = \mathcal{E}_{\text{out}}(t) + \mathcal{E}_f(t) + \mathcal{E}_m(t). \tag{4.7}$$

The corresponding normalized fractions are

$$\Phi_{\text{out}}(t) = \frac{\mathcal{E}_{\text{out}}(t)}{\mathcal{E}_{\text{in}}(t)}, \qquad \Phi_f(t) = \frac{\mathcal{E}_f(t)}{\mathcal{E}_{\text{in}}(t)}, \qquad \Phi_m(t) = \frac{\mathcal{E}_m(t)}{\mathcal{E}_{\text{in}}(t)}, \tag{4.8}$$

and obey $\Phi_{\text{out}} + \Phi_f + \Phi_m = 1$.

#### 4.3.2. Spatial moments of the thermal front

Front moments use the limit of rapid transverse equilibration and negligible longitudinal conduction. The front function $S(\xi,\tau) = \overline{\Theta}_f(L\xi, t_a\tau)$ is the aperture-averaged temperature continued on $\xi \in [0,\infty)$ under the same transport and exchange equations, driven by the velocity computed over $0 < x < L$. It decreases from one to zero and is the survival function of a front-position distribution, whose variance is nonnegative (Appendix C). Here $S(1,\tau)$ is the normalized outlet temperature.

Integration by parts gives the raw moments directly from the tail,

$$m_q(\tau) = q\int_0^\infty \xi^{q-1} S(\xi,\tau)\,\mathrm{d}\xi, \qquad q \ge 1. \tag{4.9}$$

Writing $\mathcal{M}(\tau) = m_1(\tau)$, the first two moments yield the variance, the dimensional mean front position $x_c$ and its variance $s_x^2$,

$$\mathcal{V}(\tau) = 2\int_0^\infty \xi S(\xi,\tau)\,\mathrm{d}\xi - \mathcal{M}^2(\tau), \tag{4.10}$$

$$x_c(t) = L\,\mathcal{M}(\tau), \qquad s_x^2(t) = L^2\,\mathcal{V}(\tau). \tag{4.11}$$

$\mathcal{M}$ may exceed unity; it is then the centroid of the front distribution on the half-line, not a physical position within the modeled fracture.

The moments are evaluated from limit-model identities, avoiding spatial quadrature of the upwind field (Appendix D.3). The dimensionless Laplace variable conjugate to $\tau$ is $\hat{s} = st_a$. Define the matrix-exchange symbol $\mathcal{K}(\hat{s}) = t_a/[b\,\widehat{H}_m(\hat{s}/t_a)]$, where $\widehat{H}_m$ is the chosen wall compliance at $q_x = 0$ (Section 3.1). Set $\Psi(\hat{s}) = \hat{s} + \mathcal{K}(\hat{s})$. The associated matrix-memory kernel $k$ is defined by $\hat{k}(\hat{s}) = \mathcal{K}(\hat{s})/\hat{s}$, and the fracture-residence propagator is $g = \mathscr{L}^{-1}_{\hat{s}\to\tau}[1/\Psi]$, where $\mathscr{L}^{-1}$ denotes inverse Laplace transformation. Then

$$\mathcal{M}(\tau) = \int_0^\tau g(\tau-\tau')\,U(\tau')\,\mathrm{d}\tau', \qquad m_2(\tau) = 2\int_0^\tau g(\tau-\tau')\,U(\tau')\,\mathcal{M}(\tau')\,\mathrm{d}\tau'. \tag{4.12}$$

For a unit thermal pulse entering the fracture at $\tau'$, $g(\tau-\tau')$ is the fraction in the fluid at $\tau$, including heat returned by the matrix. For finite capacity it approaches the equilibrium fluid share $(1+C_m)^{-1}$. The mean advance thus weights velocity by the mobile heat fraction.

These identities accept any positive velocity history from the nonlinear solver or reconstruction, within the stated limit model. Against extrapolated full-field moments at $\Pi_1 = 10^2$, discrepancies are below 1.6% in the mean and 4.3% in the variance; tested contrasts, window and extrapolation checks are specified in Appendix D.3. BTC and energy use the physical interval and Eq. (2.5).

Strict first-order moments are nonuniform: the variance becomes negative at late times under cooling for $\lambda \gtrsim 1/2$ (Appendix C). The finite-contrast figures therefore use the exponential velocity reconstruction of Eq. (3.11) to drive both moments through Eq. (4.12). For the same matrix closure, positive velocity histories generate admissible front distributions. Define their relative velocity error up to time $t$ by $\varepsilon_U = \sup_{0\le t'\le t} |\ln[u_{\text{rec}}(t')/u(t')]|$. Subscript rec identifies reconstructed moments; unsubscripted moments use the nonlinear-reference velocity.

The positive-kernel representation in Appendix C gives, for $\tau > 0$,

$$\left|\ln\frac{\mathcal{M}_{\text{rec}}}{\mathcal{M}}\right| \le \varepsilon_U, \qquad \left|\ln\frac{\mathcal{V}_{\text{rec}}}{\mathcal{V}}\right| \le 2\varepsilon_U. \tag{4.13}$$

For the three thicknesses of Section 5.5 at $\lambda = \pm 0.6$, $\varepsilon_U = 0.009$–$0.013$, with observed errors of 0.6%–0.9% in the mean and 1.1%–1.7% in the variance (window and refinement: Appendix D.3).

### 4.4. Model-form sensitivity

We distinguish viscosity-law sensitivity, local contrast truncation and matrix-memory choice.

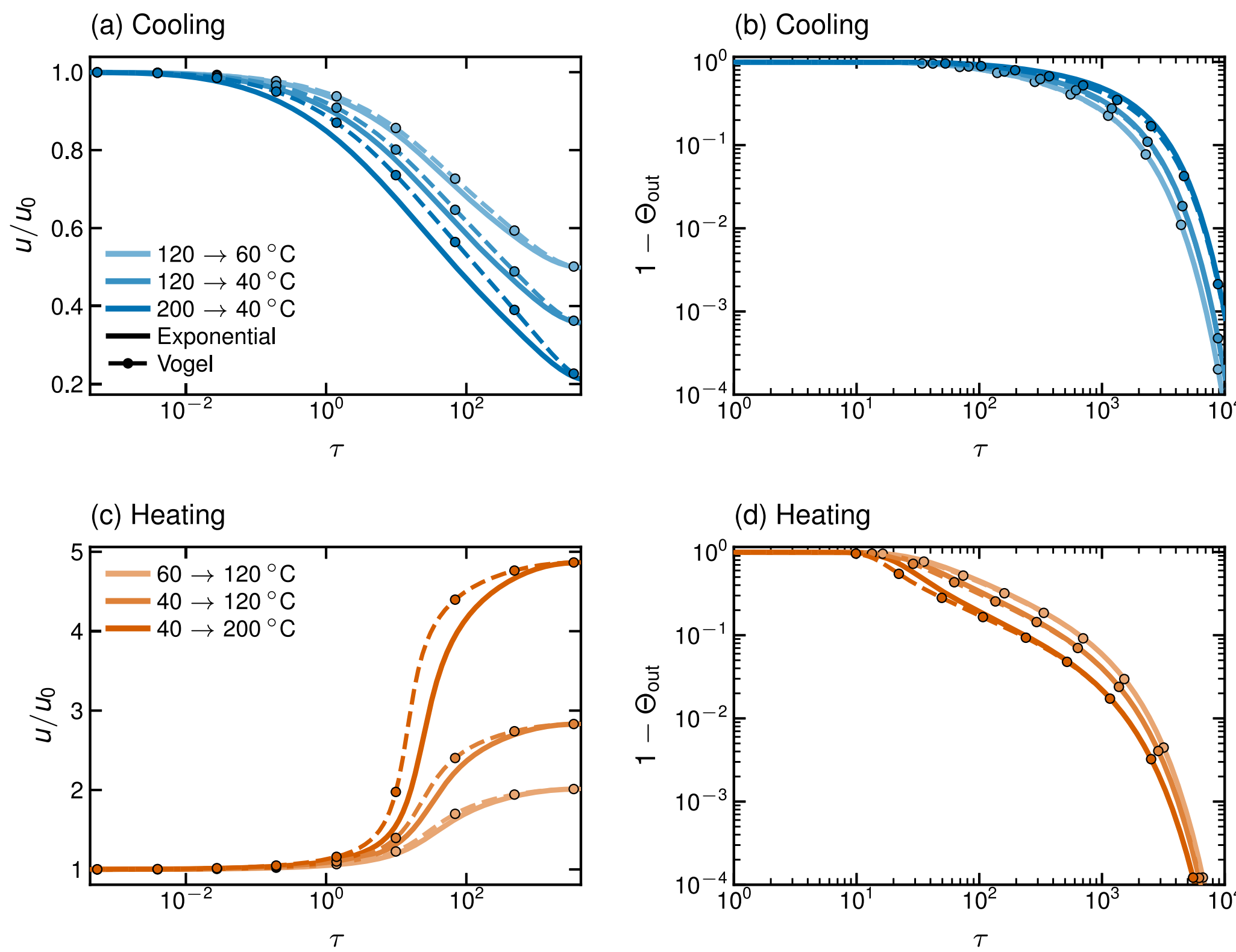

**Fig. 2.** Coupled viscosity-law sensitivity at $\Pi_1 = 103.8$ and fixed pressure. Exponential (solid) and Vogel (dashed with circles) laws match endpoint viscosities. Top: cooling; bottom: reversed heating intervals, with temperature pairs in the legends. Blue denotes cooling and vermilion heating; darker shades indicate larger contrasts. (a,c) Velocity on a logarithmic time axis. (b,d) Breakthrough survival $1 - \Theta_{\mathrm{out}}$ on logarithmic axes. Each pair shares the remaining inputs and numerical settings (Appendix D.3).

*Viscosity law.* Fig. 2 and Table 4 compare fully nonlinear exponential and Vogel solutions at $\Pi_1 = 103.8$ for three cooling intervals and their reversals, with $|\lambda| \approx 0.70$, 1.04 and 1.58. The cooling endpoints follow Okoroafor and Horne (2022), here imposed on the fixed-pressure reference geometry. Each pair shares the dimensionless controls and tempered memory, changing only the viscosity law (subscripts exp and V), so the comparison isolates constitutive sensitivity from reduction error. Along the temperature path $T = T_0 + \Theta \Delta T$, with $\Theta \in [0, 1]$, the pointwise viscosity departure is

$$E_\mu(T_0, \Delta T) = \max_{0 \le \Theta \le 1} \left| \frac{\mu_0 \exp(\lambda \Theta)}{\mu_{\mathrm{V}}(T_0 + \Theta \Delta T)} - 1 \right|. \tag{4.14}$$

The exponential law overestimates interior viscosity, $\mu_{\exp} \ge \mu_{\mathrm{V}}$, because $\ln \mu_{\mathrm{V}}$ is convex in $T$ (Fig. 3(c)). In all six coupled tests it delays half-breakthrough (Table 4); the largest shifts are 25.1% under cooling and 36.2% under heating for 200 ↔ 40 °C. Timing is more sensitive than outlet RMS, so large contrasts warrant a Vogel timing check. Exact-slab checks are reported in Appendix D.3.

At fixed rate the thermal response is isoviscous, so both laws share the temperature field and positivity of the hydraulic weights bounds the pressure sensitivity,

$$\min_T \frac{\mu_{\exp}}{\mu_{\mathrm{V}}} \le \frac{\Delta P_{\exp}}{\Delta P_{\mathrm{V}}} \le \max_T \frac{\mu_{\exp}}{\mu_{\mathrm{V}}}, \tag{4.15}$$

with applications in Section 5.8. At fixed pressure the thermal fields also differ; Table 4 reports the coupled velocity departures. Sensitivity depends on both endpoint temperatures, not on $\lambda$ alone.

*Contrast truncation.* For the first-order mobility linearization at the hydraulically weighted level $\Theta_{\mathrm{eff}} = A_0$ of Eq. (3.4), the local relative error is

$$\varepsilon_m(\lambda, \Theta_{\mathrm{eff}}) = \left| (1 - \lambda \Theta_{\mathrm{eff}}) \exp(\lambda \Theta_{\mathrm{eff}}) - 1 \right|. \tag{4.16}$$

**Table 4**
Vogel sensitivity: percentages on the stated output window. Temperature pairs are $T_0 \to T_{\mathrm{inj}}$ in °C. $E_\mu$ is defined by Eq. (4.14); $E_u = \max_t |u_{\exp}/u_{\mathrm{V}} - 1|$. Outlet RMS and maximum differences are normalized by $|T_{\mathrm{inj}} - T_0|$. The percentage shift is $\Delta\tau_{50} = 100(\tau_{50,\exp}/\tau_{50,\mathrm{V}} - 1)$, where $\tau_{50}$ is $t/t_a$ at $\Theta_{\mathrm{out}}(t) = 0.5$. Sampling, interpolation and refinement are specified in Appendix D.3.

| Temperature pair (°C) | $E_\mu$ (%) | $E_u$ (%) | Outlet RMS (%) | Outlet max (%) | $\Delta\tau_{50}$ (%) |
|---|---|---|---|---|---|
| 120 → 60 | 4.84 | 2.92 | 0.51 | 1.38 | 5.14 |
| 120 → 40 | 10.36 | 5.48 | 1.01 | 2.68 | 10.13 |
| 200 → 40 | 29.26 | 12.15 | 2.44 | 6.39 | 25.08 |
| 60 → 120 | 4.84 | 4.73 | 0.64 | 2.18 | 6.25 |
| 40 → 120 | 10.36 | 11.24 | 1.42 | 5.32 | 13.47 |
| 40 → 200 | 29.26 | 34.11 | 4.02 | 18.16 | 36.18 |

It stays below 5% for $|\lambda \Theta_{\mathrm{eff}}| \le 0.28$ (Fig. 3(d)). This local diagnostic does not bound transport or reconstruction errors; transport accuracy is assessed in Section 5.2.

*Matrix memory.* For $t > 0$ and $s > 0$, define the relative time- and Laplace-domain departures between the matrix responses as

$$\varepsilon_H(t) = \left| \frac{H_m^{\mathrm{temp}}(t)}{H_m^{\mathrm{ex}}(t)} - 1 \right|, \qquad \varepsilon_D(s) = \left| \frac{\widehat{H}_m^{\mathrm{temp}}(s)}{\widehat{H}_m^{\mathrm{ex}}(s)} - 1 \right|. \tag{4.17}$$

For the tempered closure, the time-domain wall-response kernel differs from the exact slab by up to 20% (Fig. 3(a); Appendix B). Kernel departures do not bound observable errors; closure choice affects crossover timing, tails and spreading. Timing biases are summarized in Section 3.1, and timing and heat biases are given in Table B.1; Section 5.6 explains the 50% late-variance bias.

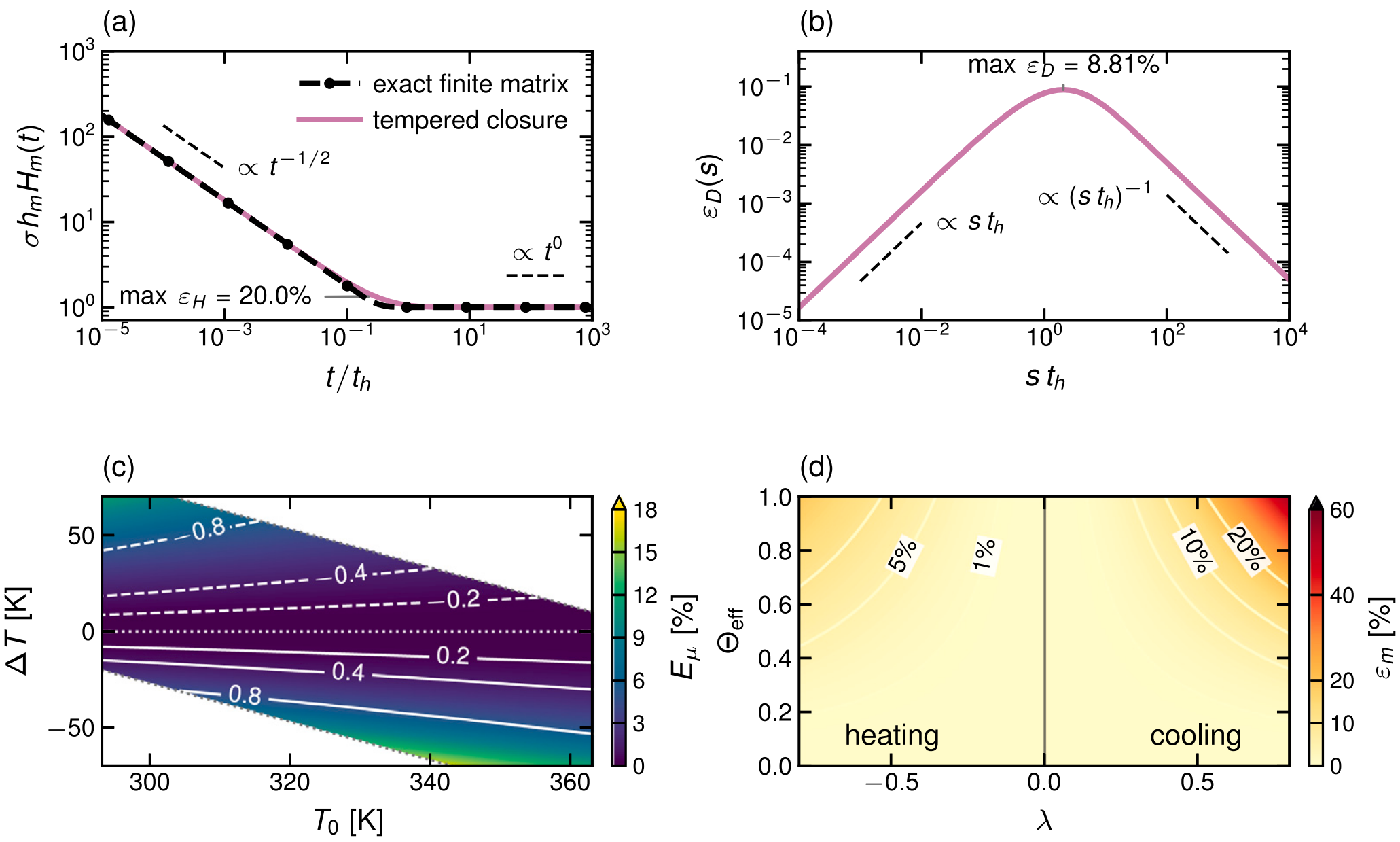

**Fig. 3.** Model-form sensitivity. (a) Tempered (solid) and exact finite-slab (dashed with circles) kernels in time, with their maximum departure $\varepsilon_H$ annotated; (b) kernel departure $\varepsilon_D$ on the positive Laplace axis. (c) Interior viscosity error $E_\mu$ of the endpoint-matched exponential law over $\Delta T$; white curves are viscosity-contrast contours. (d) Pointwise error $\varepsilon_m$ of the linearized mobility against $\lambda$ and the hydraulically weighted level $\Theta_{\mathrm{eff}}$. Panels (c,d) test constitutive forms, not a transport configuration. Kernel metrics follow Eq. (4.17).

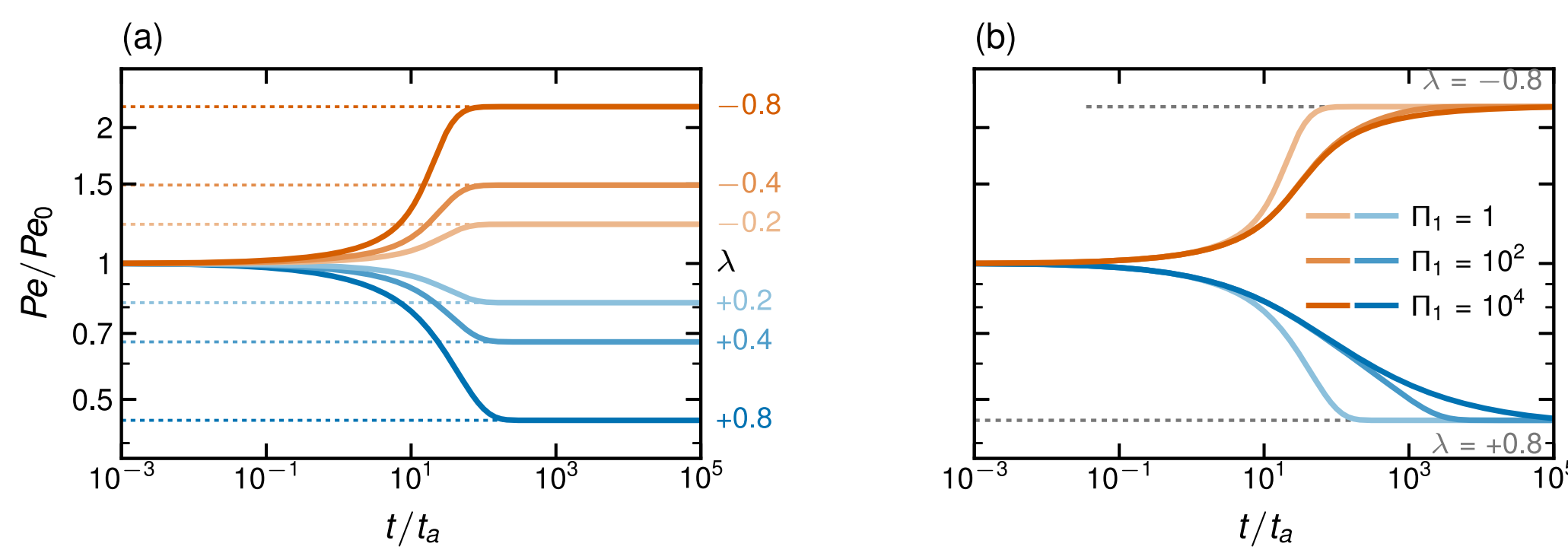

**Fig. 4.** Nonlinear hydraulic response at fixed pressure difference for the reference configuration of Table 3. (a) Viscosity-contrast dependence at $\Pi_1 = 1$. (b) Dependence on matrix timescales at $\lambda = \pm 0.8$ along $C_m = \pi \mathcal{A}\sqrt{\Pi_1}$. Dotted lines mark the exact terminal response $Pe/Pe_0 = U \to \exp(-\lambda)$; orange denotes heating and blue cooling.

Contrast truncation vanishes as $\lambda \to 0$ and differences between endpoint-matched viscosity laws vanish when the endpoint temperatures coincide, whereas matrix-closure differences can persist at $\lambda = 0$ (Sections 5.6 and 5.7).

## 5. Results

Unless stated otherwise, the results and figure captions refer to the reference configuration of Table 3 and the tempered closure; the displayed sweeps vary only $h_m$ (hence $\Pi_1$ and $C_m$ along their physical trajectory) and $\lambda$. The two closures are compared directly in Sections 5.6 and 5.7. Repeating the heating and cooling runs of Sections 5.3 and 5.5 with the exact slab preserves their ordering. Relative changes in front mean and variance differ by at most 1.4 percentage points; half-breakthrough shifts are more sensitive near the crossover, differing by up to 11 percentage points Appendix D.3.

### 5.1. Thermo-viscous modification of the hydraulic response

Viscosity contrast fixes the terminal discharge, while diffusive uptake, the matrix timescales $t_1$ and $t_r$, and the contrast itself set the approach (Section 3.4). Fig. 4 separates hydraulic amplitude from timing ($Pe(t)/Pe_0 = U$, Eq. (2.36)).

The feedback sign follows directly from the integrated resistance: as the heated portion of the fracture grows, its lower viscosity further increases discharge and accelerates subsequent advance, whereas cooling raises resistance and progressively slows the propagation that produces it. Increasing $\Pi_1$ prolongs the semi-infinite regime; beyond a threshold $\Pi_1$, further thickening no longer affects the observed response. In the semi-infinite leading-order limit, $(1-U)/\lambda = -U_1 + O(\lambda)$ (Eq. (A.5)). If the front contribution beyond the outlet is negligible, $-U_1$ approximates the isoviscous front centroid; the discharge anomaly thus measures retarded front advance through the hydraulic resistance.

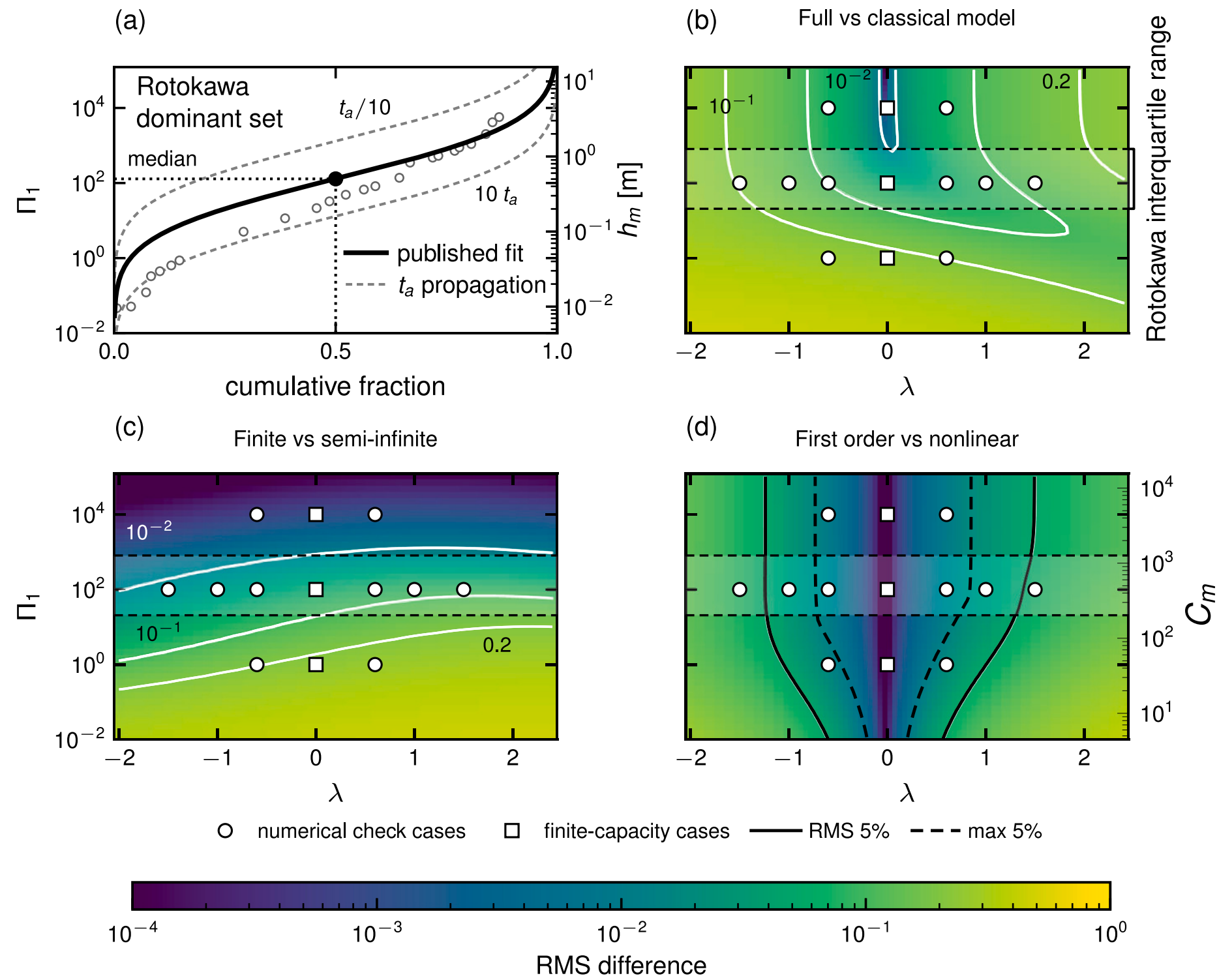

**Fig. 5.** Outlet-model comparisons and field-spacing context. (a) Rotokawa spacing cumulative distribution function (CDF) (Massiot et al., 2017a,b): empirical points and the reconstructed published lognormal curve, retained independently of the lower empirical trace. Gray curves use $t_a/10$ and $10t_a$ in the spacing projection. Its mapped interquartile range is marked by dashed lines in (b–d). The projected Rotokawa spacings indicate regime placement only; no site-calibrated simulation is shown. (b) Nonlinear finite matrix versus constant-viscosity semi-infinite transport. (c) Finite-memory discrepancy $E_s$ at matched $\lambda$. (d) First-order discrepancy $E_p$ on the same tempered closure: solid black contours mark RMS = 5%, dashed black contours sampled maximum = 5%, over the window specified in Appendix D.3. Both compare first-order and nonlinear outlet solutions, not the moment reconstruction. The common logarithmic RMS scale is $10^{-4}$–$10^{0}$; white contours in (b,c) mark equal RMS errors. Circles and squares locate numerical-check and finite-capacity cases. The maps compare models under the reference configuration, using tempered memory and the exponential viscosity law; the water-law tests reach $|\lambda| = 1.58$. Norms and sampling: Appendix D.3; spacing reconstruction: Appendix D.5.

### *5.2. Departure from the classical description and verification of the perturbative approach*

At matched $\lambda = \pm 0.6$, the finite-memory discrepancy exceeds the first-order discrepancy in the checked thinner matrices for both tempered and exact-slab memory; their ordering is reversed at $\Pi_1 \approx 10^4$ (Table 5). Over the sampled contrasts $|\lambda| \leq 1$, the first-order RMS discrepancy remains below 5% for $\Pi_1 \geq 1$. This RMS criterion does not imply uniform accuracy along the breakthrough curve: on the same window, peak outlet-temperature differences reach 23% of $|T_{\text{inj}} - T_0|$ for $|\lambda| \leq 1$ and 9% for $|\lambda| \leq 0.6$, both at $\Pi_1 \approx 1$ near breakthrough.

Fig. 5 places the response in the range of matrix timescales implied by field block sizes. It maps those sizes onto $\Pi_1$ and separates departure from the constant-viscosity, semi-infinite model (panel (b)), finite-matrix memory at matched viscosity contrast (panel (c)), and first-order truncation on the same tempered kernel (panel (d)). Denote by $E_s$ the RMS outlet discrepancy between finite and semi-infinite nonlinear models at matched $\lambda$, and by $E_p$ that between nonlinear and first-order finite-matrix models with the same tempered closure. Norms, sampling and maximum-error checks are given in Appendix D.3 (Table D.2). The classical comparison changes both storage and viscosity.

The common color scale compares the three discrepancies; its $10^{-4}$ floor is graphical, and the first-order error vanishes at $\lambda = 0$. Refinement checks are given in Appendix D.3.

At the largest thickness, the exact slab gives $E_s < 10^{-4}$, so tempering overstates the weak finite-memory departure on the sampled window.

**Table 5**
Matched-contrast outlet discrepancies $E_s$ and $E_p$ (Section 5.2) on the output window of Table D.1. Each column group uses the stated finite-matrix closure. Slab values below $10^{-4}$ are upper bounds; verification is in Appendix D.3.

| $\Pi_1$ | $\lambda$ | Tempered | | | Exact slab | | |
|---|---|---|---|---|---|---|---|
| | | $E_s$ | $E_p$ | $E_s/E_p$ | $E_s$ | $E_p$ | $E_s/E_p$ |
| 1.038 | −0.6 | 0.19272 | 0.01719 | 11.21 | 0.19165 | 0.02013 | 9.52 |
| 1.038 | +0.6 | 0.26070 | 0.01814 | 14.37 | 0.26239 | 0.02184 | 12.01 |
| 103.8 | −0.6 | 0.03145 | 0.01146 | 2.74 | 0.02854 | 0.01145 | 2.49 |
| 103.8 | +0.6 | 0.06356 | 0.01034 | 6.15 | 0.06246 | 0.01052 | 5.93 |
| 10380 | −0.6 | 0.00061 | 0.01129 | 0.05 | $<10^{-4}$ | 0.01128 | <0.01 |
| 10380 | +0.6 | 0.00116 | 0.00951 | 0.12 | $<10^{-4}$ | 0.00949 | <0.01 |

The maps describe the tempered crossover. The spacing-distribution reconstruction is specified in Appendix D.5; source temperatures and viscosity contrasts are collected in Table D.5.

Projected with the reference transit time, the Rotokawa spacing distribution (Massiot et al., 2017b) spans finite-capacity and effectively semi-infinite blocks; since $\Pi_1 \propto h_m^2/t_a$, longer transit paths shift it toward finite capacity (Fig. 5(a), gray curves). The numerical-check cases at $\Pi_1 \approx 1$ and $10^2$ lie where departure from the semi-infinite description is no longer negligible. Physically, moving across the map changes whether the outer matrix boundary is reached within the thermal observation window (Section 4.2). The first-order remainder is $O(\lambda^2)$ as $\lambda \to 0$ at fixed transport parameters, whereas departure from the constant-viscosity, semi-infinite description persists at $\lambda = 0$. Finite-capacity model departure should therefore be resolved before adding higher thermo-viscous order when crossover timing or spreading is the target.

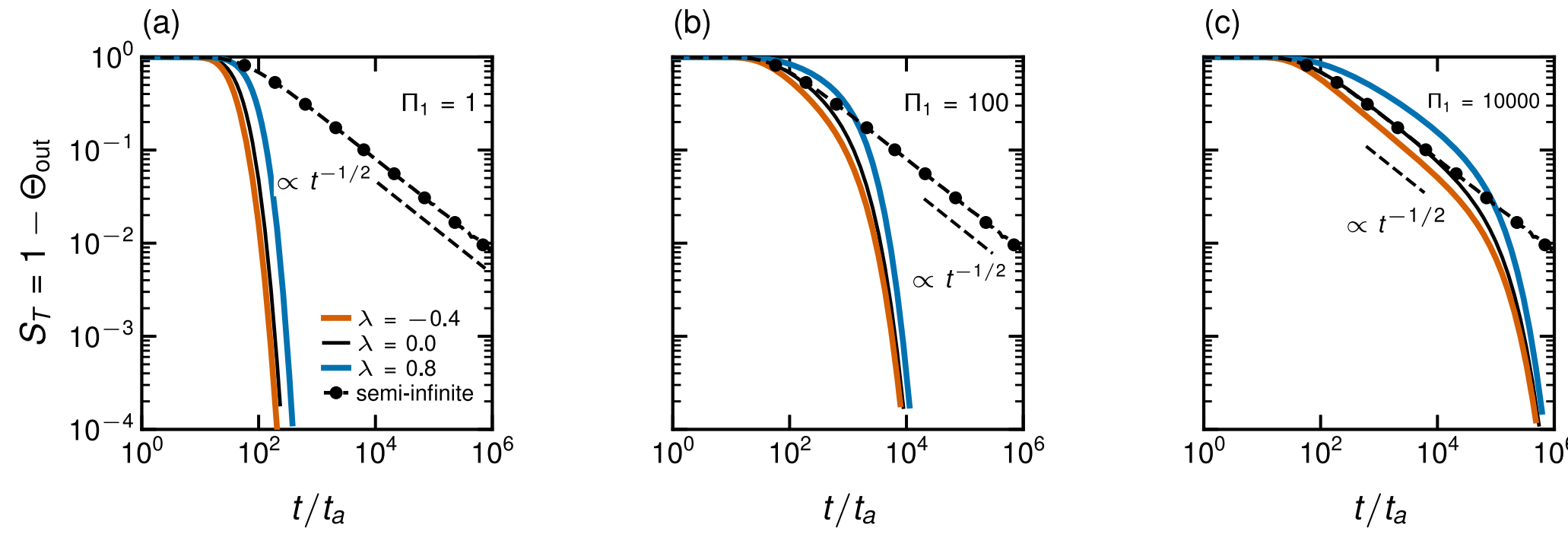

**Fig. 6.** Complementary outlet breakthrough for the reference configuration of Table 3: $S_T$ for increasing $\Pi_1$ (left to right) and heating, isoviscous, and cooling conditions ($\lambda = -0.4$, 0, 0.8; orange, black, and blue). Solid curves use the tempered closure; dashed curves with circles are the independent isoviscous semi-infinite reference of the Roubinet et al. (2012) class; dashed guides mark the semi-infinite $t^{-1/2}$ memory regime, which the tempered response follows only before the cutoff $t_h$.

### 5.3. Thermal breakthrough under viscosity feedback

Heating advances every outlet percentile, cooling delays it, and per unit contrast the two shifts are nearly log-symmetric. The residual branch asymmetry is the higher-order feedback shown in Fig. 4.

Fig. 6 shows the complementary outlet temperature $S_T = 1 - \Theta_{out}$ for $\Pi_1 = 1$, $10^2$, and $10^4$.

Despite these large timing shifts, the curves remain approximately parallel during the matrix-controlled part of breakthrough and retain

$$1 - \Theta_{out} \propto t^{-1/2}, \tag{5.1}$$

because viscosity changes the advection timescale, not the transverse diffusion operator.

Finite thickness sets the timescale of this memory regime: the response leaves the $t^{-1/2}$ law once the matrix capacity is felt, on the tempered cutoff $t_h$ of Eq. (3.1); the observable-specific exact-slab thresholds are discussed in Appendix B. Across the panels the cutoff moves with the matrix timescale: at $\Pi_1 = 1$ the cutoff $t_h \approx 10 t_a$ precedes outlet breakthrough, so the outlet shows no $t^{-1/2}$ segment and falls steeply below the semi-infinite reference, whereas at $\Pi_1 = 10^4$ the two remain nearly indistinguishable over most of the displayed window.

### 5.4. Energy partition and matrix thermal storage

The matrix dominates the transient energy budget. Fig. 7 partitions cumulative thermal input among outlet discharge, fracture storage, and matrix storage.

In the early advective loading regime before breakthrough, the heated length grows as $t$ and matrix penetration as $t^{1/2}$, giving $\Phi_m \propto t^{1/2}$. During the subsequent diffusion-controlled advance, fracture storage grows as $t^{1/2}$, giving $\Phi_f \propto t^{-1/2}$. After breakthrough, $\Phi_f \propto t^{-1}$ while a locally semi-infinite matrix gives $\Phi_m \propto t^{-1/2}$; slab saturation changes the latter to $t^{-1}$. Viscosity shifts these regimes in time without changing their exponents. The limit $\Phi_{out} \to 1$ reflects cumulative-input normalization, not loss of stored heat.

### 5.5. Thermo-viscous front kinematics

In the early matrix-exchange regime, all cases follow the semi-infinite scalings, $\mathcal{M} \propto t^{1/2}$ and $\mathcal{V} \propto t$. Finite capacity steepens the mean toward $\mathcal{M} \propto t$; at $\Pi_1 = 10^4$ this late regime lies beyond the plotted window. Fig. 8 follows the thickness trajectory of Eq. (4.1).

When the isoviscous centroid reaches the outlet, $\mathcal{M}(\lambda = 0) = 1$, heating advances and broadens the front, whereas cooling retards and sharpens it. This ordering persists across the displayed matrix timescales. The change in front width is not an additional dispersive coefficient. The matrix-exchange delays are set in physical time and projected onto the fracture by the velocity, so a stationary velocity change rescales the whole front, $\mathcal{M} \propto U$ and $\mathcal{V} \propto U^2$ (Appendix C); the evolving velocity history modulates this rescaling. The factor-of-two sensitivity is exact for a stationary rescaling; the time-dependent error bounds are given in Eq. (4.13). Shear-induced dispersion remains outside the model (Section 2.1). The exponential reduced model of Eq. (3.11) follows both moments over the whole window, with the accuracy quoted in Section 4.3.2. The finite-capacity departures already visible in the $\Pi_1 = 1$ column are isolated at $\lambda = 0$ in Section 5.6.

### 5.6. Finite-matrix control of front advancement and spreading

At $\lambda = 0$, finite matrix thickness sets the crossover time, while the relaxation-time distribution controls late spreading. Fig. 9 isolates finite capacity at $\lambda = 0$. The semi-infinite memory is $k_L(\tau) = \mathcal{A}/\sqrt{\pi\tau}$. The variance closure bias is $E_{\mathcal{V}} = \mathcal{V}_t/\mathcal{V}_e - 1$, while $\delta_{\mathcal{V},j} = \mathcal{V}_j/\mathcal{V}_L - 1$ measures the departure of closure $j \in \{e, t\}$ from the semi-infinite variance.

Departure from the common semi-infinite scalings occurs at $\tau/\Pi_1 = O(1)$, confirming the $h_m^2$ timescale: the crossover begins once the thermal penetration depth $\sqrt{\kappa_{Tm} t}$ becomes an $O(1)$ fraction of $h_m$ ($t \sim t_1$) and the blocks begin to feel the no-flux outer boundary; thereafter the remaining matrix volume is progressively filled, and the fracture and matrix approach a capacity-retarded state. Thickness therefore controls the position of the front: both closures approach the same late mean slope, $\mathcal{M} \sim \tau/(1 + C_m)$ (panel (a)).

The variance responds differently (panel (b)). Before the slab is felt it grows as the semi-infinite law $\mathcal{V}_L = (2 - 4/\pi)\tau/\mathcal{A}^2$ (Eq. (B.17)), and after filling it returns to Fickian growth with

$$\mathcal{V} \sim \frac{2 C_m \bar{\tau}_m \tau}{(1 + C_m)^3} = \frac{2c\,\tau}{\mathcal{A}^2}\left(\frac{C_m}{1 + C_m}\right)^3, \qquad \bar{\tau}_m = c\,\pi^2 \Pi_1, \tag{5.2}$$

where $\bar{\tau}_m$ is the capacity-weighted mean residence time in the matrix, in units of $t_a$, $c = 1/3$ for the exact slab and $c = 1/2$ for the tempered closure. A thicker slab holds heat longer ($\bar{\tau}_m \propto h_m^2$) but retards the front more ($C_m \propto h_m$), and the two effects cancel: at fixed $\mathcal{A}$ the late variance is nearly independent of thickness, which sets only the time of the transition. The sign of its departure from the semi-infinite law, however, depends on the closure. For the three thicknesses $\Pi_1 = 1, 10^2, 10^4$, the exact slab lowers the late variance by 8%–14%, whereas tempering raises it by 29%–38%. The tempered-to-slab ratio is 3/2: a 50% late-variance bias, consistent with Eq. (B.24) in the large-$C_m$ limit.

Panel (c) compares the memory responses associated with the residence times in Eq. (5.2); the memory spectra are detailed in Appendix

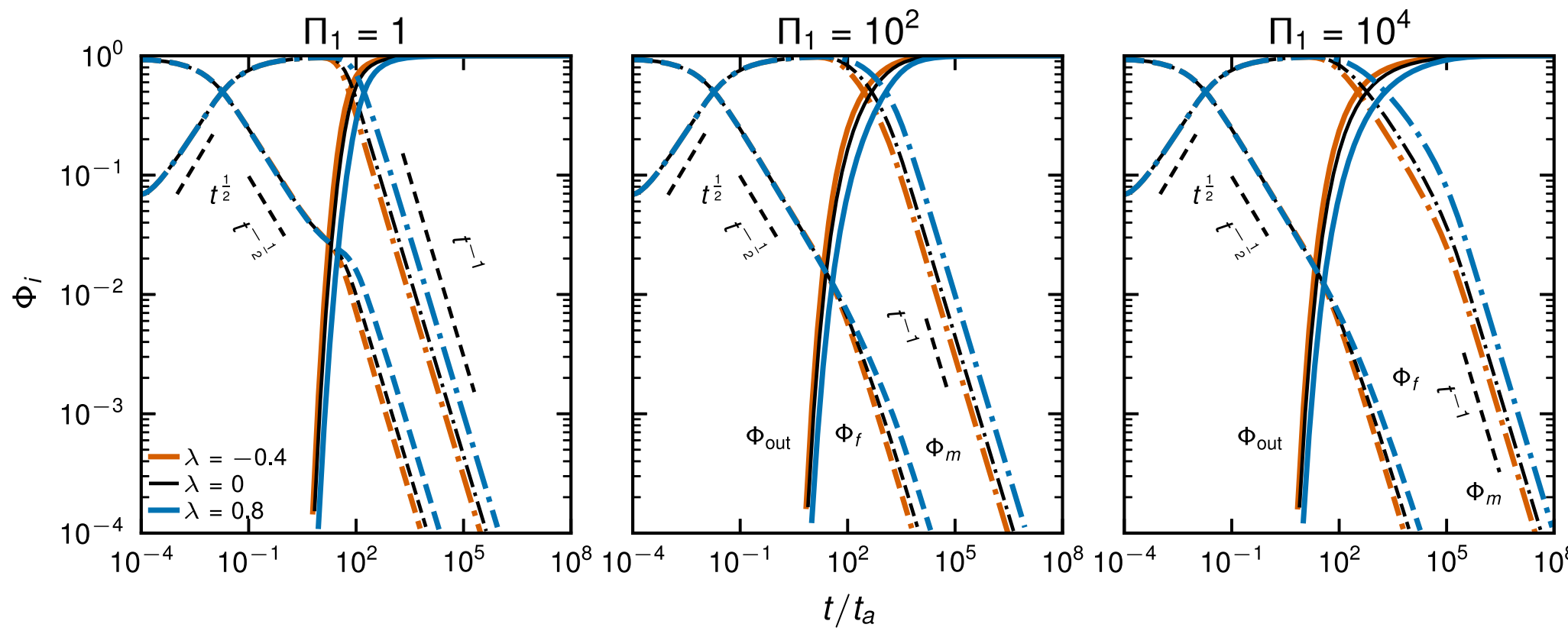

**Fig. 7.** Cumulative energy partition for the reference configuration of Table 3, computed with the tempered closure for $\Pi_1 = 1,\ 10^2,\ 10^4$ (left to right). Solid, dashed, and dash-dotted curves denote the fractions $\Phi_i$, $i \in \{\mathrm{out}, f, m\}$, assigned to outlet discharge, fracture storage, and matrix storage; orange, black, and blue denote heating, isoviscous, and cooling conditions ($\lambda = -0.4,\ 0,\ 0.8$). Guides mark matrix loading ($t^{1/2}$), fracture-storage decay before breakthrough ($t^{-1/2}$), and late decay ($t^{-1}$).

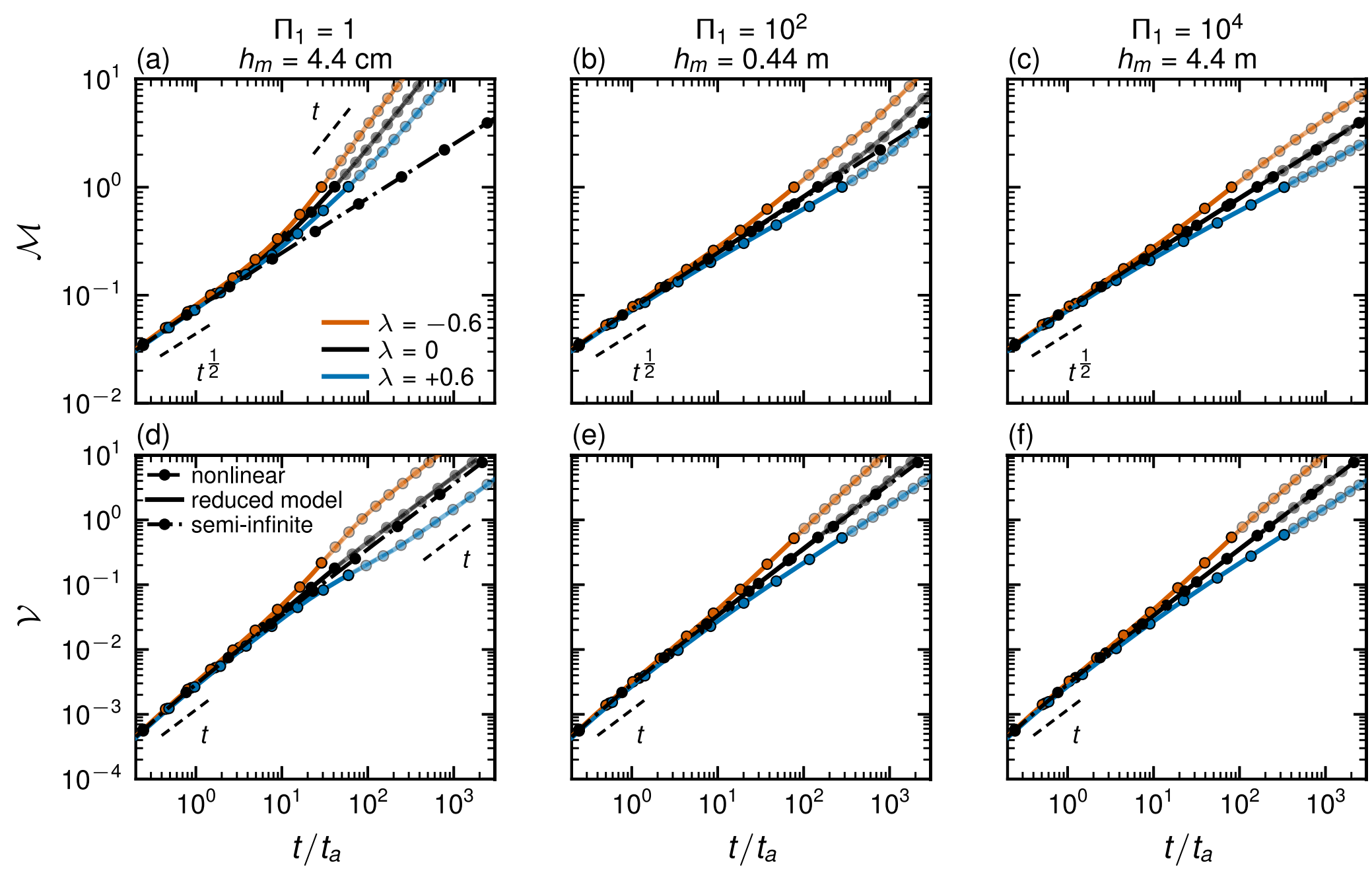

**Fig. 8.** Thermal-front mean $\mathcal{M}$ (a–c) and variance $\mathcal{V}$ (d–f) for the reference configuration of Table 3, at $\Pi_1 = 1, 10^2, 10^4$ (left to right). Orange, black, and blue denote $\lambda = -0.6, 0, 0.6$; solid curves are the exponential reduced model of Eq. (3.11), dashed curves with circles the nonlinear solution, and dash-dotted curves with circles the semi-infinite reference. All moments are evaluated from the moment representation of Eq. (4.12) with the corresponding velocity history. Black dashed guides mark the local scaling regimes; faint segments of the finite-matrix curves lie beyond $x_c = L$.

B. The effective memory captures the diffusion-to-capacity crossover; for the planar geometry, quantitative onset thresholds, breakthrough-tail fits and front spreading use the exact slab.

### 5.7. Finite-matrix similarity and closure fidelity

Under strong matrix uptake, the isoviscous field collapses onto a two-variable similarity surface. For either finite closure, define the relative departure from the Lauwerier mean $\mathcal{M}_L$ of Eq. (A.3) by $\delta_m = (\mathcal{M} - \mathcal{M}_L)/\mathcal{M}_L$. For the tempered closure and a prescribed relative departure $\delta > 0$, the small-departure law is

$$\delta_m(\tau) \sim \frac{\tau}{3\pi^2 \Pi_1}, \qquad \frac{\tau_\delta}{\Pi_1} \sim 3\pi^2 \delta, \tag{5.3}$$

as $\tau/\Pi_1 \to 0$ and $\delta \to 0$, respectively, where $\tau_\delta$ is the time at which the tempered mean reaches $\delta_m = \delta$. This predicts the tempered onset under strong uptake, not the exact-slab prefactor. In the strong-uptake limit the aperture-averaged field depends on position and time only through $\eta = \mathcal{A}\xi/\sqrt{\Pi_1}$ and $r = t/t_1 = \tau/\Pi_1$: $\eta$ is the capacity delay $C_m x/u_0$ in units of $\pi t_1$, and $r$ is the elapsed time in units of $t_1$. For closure $j$, the scaled front is

$$\mathcal{T}_j(\eta, r) = S_j\left(\frac{\sqrt{\Pi_1}}{\mathcal{A}}\eta, \Pi_1 r\right). \tag{5.4}$$

Fig. 10 compares these master fields; Appendix B derives their similarity laws. The exact-slab field is the parallel-fracture solution of Gringarten et al. (1975) without fracture storage, written in these coordinates; the tempered field, the moment laws and the closure comparison extend it.

The leading-order field is a single surface in $(\eta, r)$; BTC collapse applies at matched $\eta$, not automatically at a common physical outlet. The magnitude of the signed difference mapped in panel (b) never exceeds 0.07; the difference changes sign across the front: the closures disagree about the position and width of the front, not about the amplitude of the field. The closures share total capacity and the late-time slope of the mean, so the relative difference between their finite-capacity means decays after the crossover. Their departure from the semi-infinite mean, shown in panel (e), grows instead. The variance retains the different distributions of matrix relaxation times and hence the finite bias of panel (f) (Section 5.6; Appendix B).

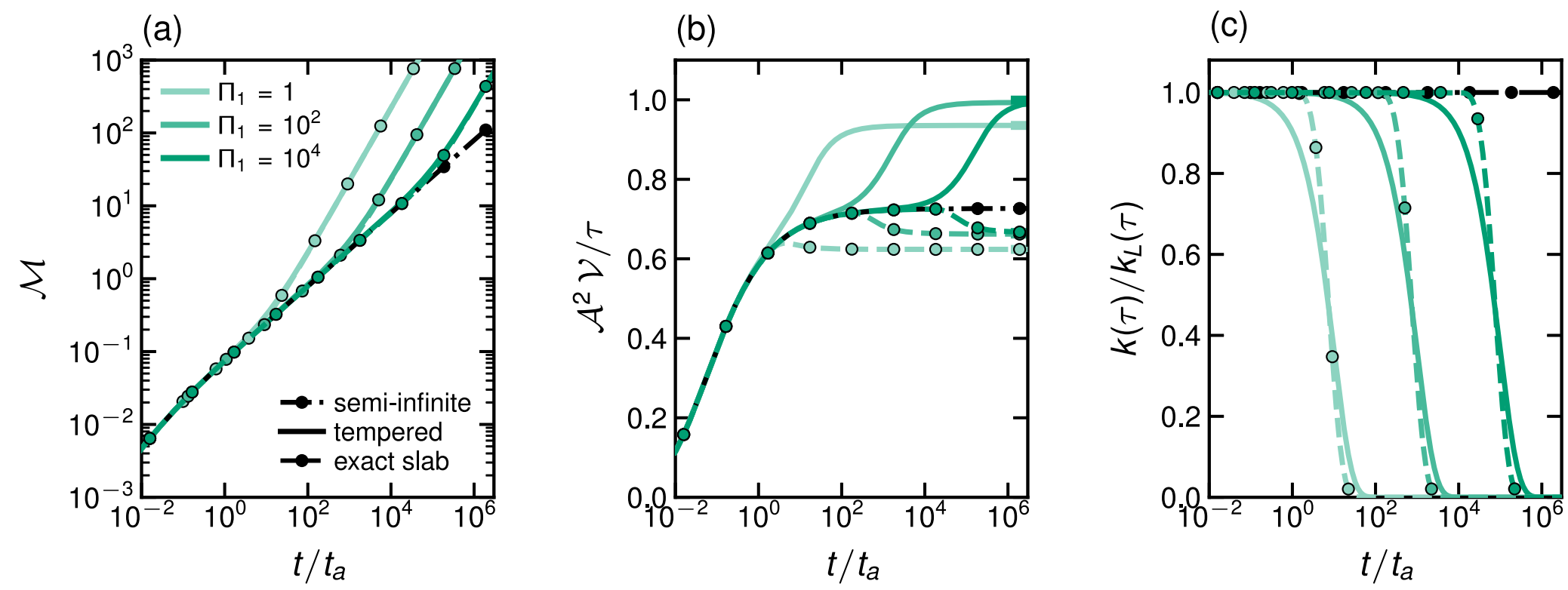

**Fig. 9.** Effect of slab thickness at $\lambda = 0$ and fixed exchange amplitude $\mathcal{A}$, for the reference configuration of Table 3 at $\Pi_1 = 1, 10^2, 10^4$ (the columns of Fig. 8); darker green denotes a thicker slab. (a) Front mean. (b) Compensated variance $\mathcal{A}^2\mathcal{V}/\tau$; end ticks mark the late-time limits of Eq. (5.2). (c) Matrix memory function relative to the semi-infinite one, $k/k_L$. Solid curves are tempered; exact-slab curves are dashed with circles, and the semi-infinite limit is dash-dotted with circles. Moments are evaluated from Eq. (4.12) with $U = 1$.

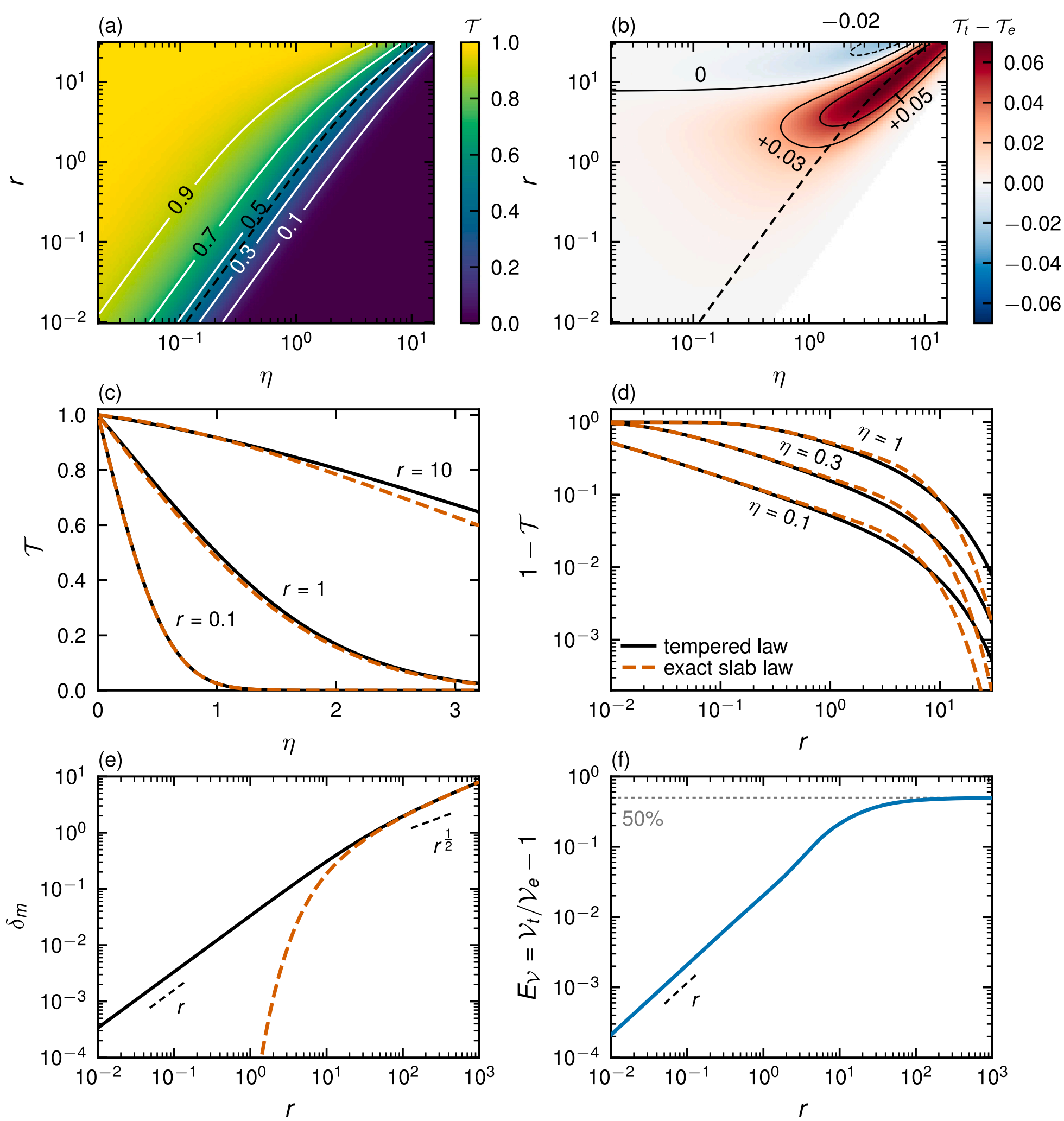

**Fig. 10.** Finite-matrix similarity at $\lambda = 0$ in the strong-uptake limit. (a) Tempered master field $\mathcal{T}_t$ in $(\eta, r)$. (b) Signed field difference $\mathcal{T}_t - \mathcal{T}_e$. Black dashed curves in (a,b) trace the tempered centroid. (c,d) Theoretical profile and breakthrough cuts at matched similarity coordinates. (e) Relative mean departure $\delta_m$ from the semi-infinite limit for each closure. (f) Variance bias $E_\mathcal{V} = \mathcal{V}_t/\mathcal{V}_e - 1$. Black solid and vermilion dashed curves denote the tempered and exact-slab laws; black dashed guides indicate asymptotic slopes and the blue curve gives the variance ratio minus one.

### 5.8. Fixed-rate protocol and the pressure observable

At fixed rate in a one-dimensional uniform-aperture channel, viscosity changes the required pressure while the thermal response remains isoviscous, so the coupling is one-way. This section tests that formulation against a published numerical benchmark; a pressure-response parameter survey is left to future work.

At fixed rate, with the cross-sectionally averaged velocity $u_q$ imposed, Eq. (2.17) gives the dual resistance relation

$$\frac{\Delta P(t)}{\Delta P_0} = \frac{1}{L}\int_0^L \frac{\mathrm{d}x}{m_H(x,t)}, \qquad \Delta P_0 = \frac{3\mu_0 u_q L}{b^2}, \tag{5.5}$$

with $m_H$ evaluated from the prescribed-velocity thermal solution. Here $\Delta P$ is the viscous drop along the fracture, excluding well losses and gravitational circulation. The benchmark fracture is horizontal (Okoroafor and Horne, 2022); for inclined paths the gravity criterion of Appendix D.3 applies. With the benchmark inputs of Appendix D.5, this drop is about 0.5 kPa at 200 °C and rises fivefold after full cooling to 40 °C, so its detection requires separating the fracture contribution from well losses; the pressure response is a model prediction, not compared with benchmark pressures. The uniform-aperture simulations of Okoroafor and Horne (2022) prescribe the rate in a two-dimensional converging–diverging well doublet; we take the benchmark's own transit time, $1.25 \times 10^5$ s, as $t_a$. With the semi-infinite leading-order response of Eq. (A.2), we use one effective exchange amplitude $\mathcal{A}_{\mathrm{eff}} = 85.07$, fitted jointly to the four digitized outlet histories over $\tau > 1.05$ (normalized RMS 0.0038, coefficient of determination $R^2 = 0.98$; Fig. 11). Nonlinear least squares gives a nominal 95% confidence interval of 84.98–85.16 for $\mathcal{A}_{\mathrm{eff}}$ under the assumption of independent residuals. Since the residuals of each digitized curve are serially correlated, this interval should be interpreted as conditional on that assumption. Separate fits to the four histories span 83.6–86.5, below the nominal values of Appendix D.5. The two injection temperatures give the same amplitude to within 0.3%, as expected from the isoviscous fixed-rate thermal response, whereas the two reservoir temperatures differ by about 3%.

The nominal amplitude follows from aperture, transit time and rock thermal properties through Eq. (2.35), using hydraulic measurements and laboratory samples. Appendix D.5 documents the benchmark inputs and nominal values. The effective amplitude accommodates distributed doublet residence times and temperature-dependent density absent from the one-dimensional, constant-property model; the comparison tests the response shape.

Because the fixed-rate thermal response is isoviscous, the thickness results of Figs. 9 and 10 apply unchanged at fixed rate. The pressure prediction uses the endpoint-matched exponential law. Replacing it with Vogel on the same calibrated thermal field lowers $\Delta P$ by up to 15.8% for 200 → 40 °C and 10.8% for 200 → 60 °C over 0.02–5.2 years. These reductions satisfy Eq. (4.15); Appendix D.5 reports the quadrature and bounds. Resistance responds before outlet breakthrough: at five years the inset of panel (c) shows that the factor $\exp\left(\lambda\overline{\Theta}_f^{(0)}\right)$ is elevated only over the invaded near-inlet region. Channeling, aperture variability and two-dimensional in-plane flow paths would redistribute flow, breaking this decoupling and reintroducing $\lambda$ into thermal transport (Hoffmann et al., 2022; Liu et al., 2024).

## 6. Discussion

The tempered effective memory preserves early diffusion and total storage, whereas the planar-slab comparison at $\lambda = 0$ quantifies the effect of the relaxation-time distribution on breakthrough timing, tails and front spreading: the slab's late variance is two-thirds of the tempered value. These distinctions guide closure choice: tempering for elementary crossover laws, the slab for quantitative timing, tails and spreading in parameter inference (Sections 5.6 and 5.7).

The tested accuracy ranges of the reduced formulations are summarized in Table 2. The maps separately identify finite-capacity effects and perturbative truncation, which require different model refinements. Constitutive sensitivity affects breakthrough timing more than outlet RMS; quantitative timing at large temperature contrasts can use Vogel in the nonlinear solver (Section 4.4, Fig. 2).

This reduction complements three multiphysics studies with distinct objectives. Gao et al. (2026) assess hydrogen-storage performance and economics with coupled one-dimensional wellbore and three-dimensional reservoir flow, heat, mechanics, diffusion and dissolution. State-dependent viscosity enters multiphase mobility, and rock heat storage is resolved through the reservoir energy equation rather than an isolated matrix slab. Ji et al. (2026a) study how matrix reactions affect long-term heat extraction: viscosity enters Darcy mobility, while porous-matrix heat storage and advective–conductive interface fluxes are resolved explicitly. Their stimulated volume is an equivalent porous medium; reactions change aperture and permeability, while mechanics is omitted. Ji et al. (2026b) track reaction-driven morphology at fracture intersections using Navier–Stokes flow and temperature- and pressure-dependent water properties. Their isothermal walls supply heat continuously, rather than equilibrating through finite matrix storage. Here conductive memory and integrated resistance isolate finite storage and viscosity feedback at fixed geometry. The present reduced model could supply a thermal exchange component for locally conductive matrix blocks in the reservoir models, or replace prescribed wall temperatures in an intersection model. Coupling this component to matrix advection, multiphase flow and evolving geometry requires further development.

The physical assumptions delimit the use of this baseline. A fixed planar aperture and quasi-steady flow isolate thermal changes in resistance; roughness, deformation and reactions require geometry and exchange updates (Klepikova et al., 2021; De Simone et al., 2023; Lenci et al., 2026b). Thermoelastic contraction under cooling opens the aperture and opposes the viscosity increase (Guo et al., 2016); it is negligible while the relative aperture change remains well below $|\lambda|/3$. Section-mean advection requires rapid transverse diffusion; its error at the computed Graetz numbers is quantified in Section 4.1. Shear spreading at larger $Gz$ and lateral fingering need additional flow closures; heating, in which less viscous fluid displaces more viscous fluid, is the potentially unstable case, for which the laterally uniform solution is a base state. The conductive insulated matrix equilibrates, unlike a permanently cooled wall (Wylie and Lister, 1995; Helfrich, 1995). Constant fluid properties isolate viscosity: density changes by 1.4% over 55–80 °C at 0.1 MPa but by 12.6% over 40–200 °C at 10 MPa, so at large contrasts negligible buoyancy requires sub-horizontal fractures or strong pressure forcing. The criterion uses the rescaled pressure of Section 4.1 (Appendix D.3).

Applications include breakthrough prediction and interpretation of pressure histories at prescribed rate. Inputs are initial and injection temperatures, flow and pressure histories, and constraints on aperture, path length and matrix thermal properties. Spatial profiles can be obtained from borehole temperature arrays or distributed fiber-optic sensing (Hawkins et al., 2017), with sufficient coverage of a constrained flow path to estimate front moments while the front remains within the monitored path. For a chosen geometry and closure, mean and variance can provide complementary constraints on accessible storage and exchange strength. A single outlet curve constrains thickness and diffusivity jointly for a chosen geometry and closure; in the parallel-fracture interpretation, the inferred half-thickness gives center-to-center spacing $2(b + h_m)$.

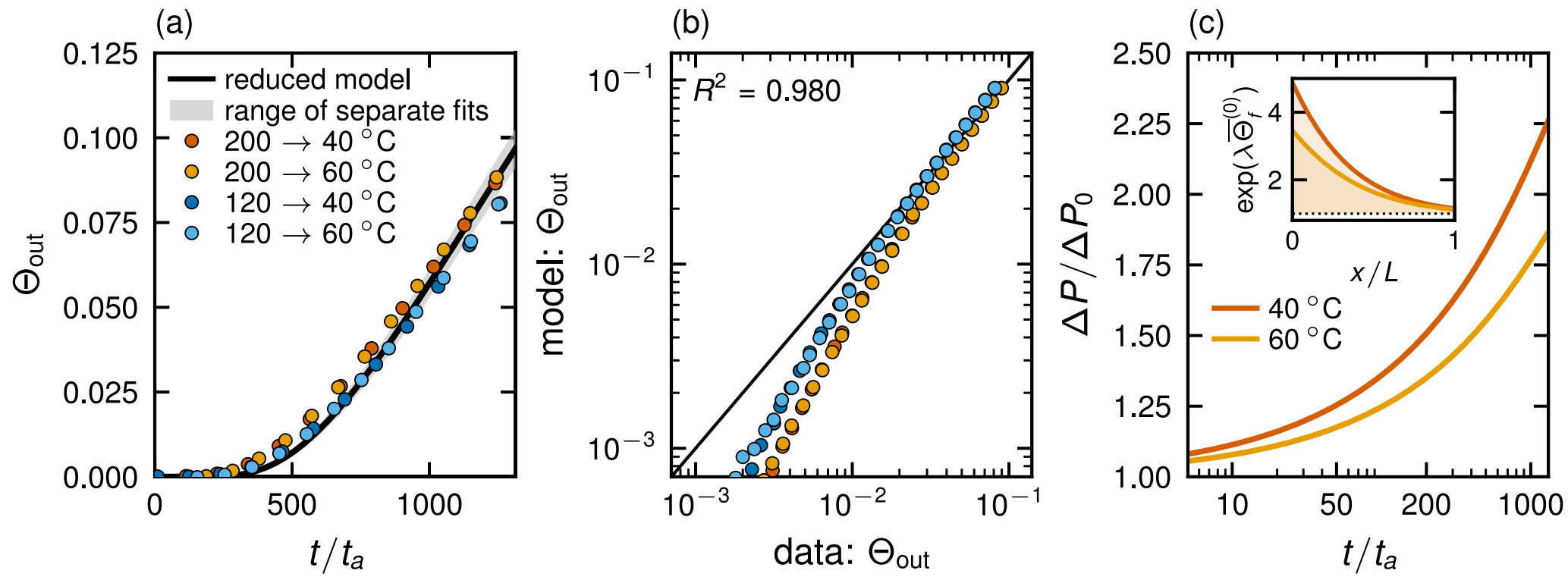

**Fig. 11.** Fixed-rate comparison with the uniform-aperture simulations of Okoroafor and Horne (2022), using one fitted $\mathcal{A}_{\mathrm{eff}} = 85.07$ to accommodate geometry and model differences. Time is normalized by $t_a = 1.25 \times 10^5$ s; legend pairs give $T_0 \to T_{\mathrm{inj}}$ in °C. (a) Digitized outlet temperatures and leading-order response; the shaded band spans the separate fits to the four histories, $83.6 \le \mathcal{A}_{\mathrm{eff}} \le 86.5$, as a sensitivity band without probabilistic meaning (Section 5.8). (b) Parity plot. The temperature-pair colors in (a) apply throughout. (c) Conditional pressure response from Eq. (5.5) for 200 → 40 and 200 → 60 °C (vermilion and orange). Solid colored curves give viscous pressure; the inset shows $\exp\left(\lambda \overline{\Theta}_f^{(0)}\right)$ at five years, with the dotted unit line marking the initial resistance. Vogel endpoint contrasts affect pressure only; benchmark inputs are given in Appendix D.5.

Integration into reservoir simulators requires conservative interface fluxes, initialized memory states and the resistance derivative in the flow solver. Aperture or property changes require updating the memory operator; single-fracture conservation benchmarks should precede network use.

The evidence comprises analytical and numerical verification, the calibrated numerical benchmark of Okoroafor and Horne (2022) for response shape, and field spacing data and literature contrasts for physical context. Field validation and network upscaling (Lenci et al., 2026a) are subsequent steps.

## 7. Conclusions

This work analyzed heat transport in a planar fracture bounded by matrix blocks of finite thermal capacity, with the fracture discharge set by the temperature-dependent hydraulic resistance. The integrated resistance closes the feedback between heat transport and flow. At fixed pressure difference, heating accelerates the flow and advances and broadens the thermal front, whereas cooling retards and sharpens it. The terminal discharge depends only on the endpoint viscosity ratio, $u_\infty / u_0 = \exp(-\lambda)$, while diffusive uptake, matrix capacity and the contrast set the time over which it is approached. At fixed rate in a uniform-aperture fracture, the thermal response is isoviscous and viscosity acts only on the fracture pressure drop, which can respond before outlet breakthrough because the resistance samples the invaded part of the fracture.

Finite matrix thickness introduces a thickness-squared timescale and controls the transition from semi-infinite uptake to capacity-limited equilibration. Under strong uptake, the isoviscous response collapses onto a two-variable similarity surface, and a tempered memory fixed by initial uptake and total capacity yields elementary crossover laws.

The perturbation hierarchy separates matrix storage from the hydraulic response it induces. The exponential velocity reconstruction preserves the exact terminal discharge and bounds front-moment errors within the front-moment limit model; its tested accuracy delimits where the reduced model can replace the nonlinear solution, which remains the reference at large contrasts. The model thus provides a single-fracture baseline for breakthrough prediction, pressure interpretation and thermal-exchange components of reservoir simulators.

## Declaration of Generative AI and AI-assisted technologies in the writing process

During the preparation of this work, the author used ChatGPT and Codex (OpenAI) and Claude (Anthropic) to assist with language refinement, manuscript organization, and consistency checks. After using these tools, the author reviewed and edited the content as needed and takes full responsibility for the content of the publication.

## Funding

A.L. acknowledges funding from the European Union's Horizon Europe research and innovation programme under the Marie Skłodowska-Curie grant agreement No. 101111216, Project GEONEAT, "Complex Fluids in Fractured Geological Media for Enhanced Heat Transfer". Views and opinions expressed are, however, those of the author only and do not necessarily reflect those of the European Union or the European Research Executive Agency (REA). Neither the European Union nor the granting authority can be held responsible for them. The funding source had no role in the study design, data collection, analysis or interpretation, manuscript preparation, or the decision to submit the article for publication.

## Declaration of competing interest

The authors declare that they have no known competing financial interests or personal relationships that could have appeared to influence the work reported in this paper.

## Appendix A. Closed-form semi-infinite limit: the thermo-viscous extension of the Lauwerier solution

While conductive penetration remains small compared with the slab thickness, matrix exchange reduces to the semi-infinite half-order operator. With rapid transverse equilibration ($Gz \ll 1$) and negligible in-fluid longitudinal conduction, the leading-order semi-infinite front $S_L^{(0)}$ obeys, in the dimensionless coordinates of Section 2.5,

$$\frac{\partial S_L^{(0)}}{\partial \tau} + \frac{\partial S_L^{(0)}}{\partial \xi} = -\mathcal{A} \frac{\partial^{1/2} S_L^{(0)}}{\partial \tau^{1/2}}, \tag{A.1}$$

with $S_L^{(0)}(\xi, 0) = 0$ and $S_L^{(0)}(0, \tau) = 1$ for $\tau > 0$. The causal half-order derivative has Laplace symbol $\sqrt{\hat{s}}$ for zero prehistory. This is the classical problem of Lauwerier (1955), transposed from the confined porous layer to the fracture–matrix geometry (Bodvarsson, 1969; Gringarten et al., 1975). Its Laplace-space solution and time-domain inversion are

$$\hat{S}_L^{(0)}(\xi,\hat{s}) = \frac{1}{\hat{s}}\exp\left(-\xi\left(\hat{s}+\mathcal{A}\sqrt{\hat{s}}\right)\right),$$
$$S_L^{(0)}(\xi,\tau) = \mathrm{erfc}\left[\frac{\mathcal{A}\,\xi}{2\sqrt{\tau-\xi}}\right]\mathrm{H}(\tau-\xi), \tag{A.2}$$

where erfc is the complementary error function and H the Heaviside step. The breakthrough curve follows by evaluating this solution at $\xi = 1$ (Eq. (4.3)).

The front moments of Section 4.3.2 have the Laplace transforms $\widehat{\mathcal{M}}_L = [\hat{s}(\hat{s}+\mathcal{A}\sqrt{\hat{s}})]^{-1}$ and $\hat{m}_{2,L} = 2[\hat{s}(\hat{s}+\mathcal{A}\sqrt{\hat{s}})^2]^{-1}$, whose inversions read

$$\mathcal{M}_L(\tau) = \frac{2}{\mathcal{A}}\sqrt{\frac{\tau}{\pi}} - \frac{1}{\mathcal{A}^2}\left[1-\exp(\mathcal{A}^2\tau)\,\mathrm{erfc}\left(\mathcal{A}\sqrt{\tau}\right)\right], \tag{A.3}$$

$$m_{2,L}(\tau) = 2\left[\frac{\tau}{\mathcal{A}^2} - \frac{6}{\mathcal{A}^3}\sqrt{\frac{\tau}{\pi}} + \frac{3}{\mathcal{A}^4} + \frac{2\mathcal{A}^2\tau-3}{\mathcal{A}^4}\exp(\mathcal{A}^2\tau)\,\mathrm{erfc}\left(\mathcal{A}\sqrt{\tau}\right)\right]. \tag{A.4}$$

In the exchange regime $\mathcal{A}^2\tau \gg 1$, these limit to $\mathcal{M}_L \sim (2/\mathcal{A})\sqrt{\tau/\pi}$ with $m_{2,L}/\mathcal{M}_L^2 \to \pi/2$.

The closed forms hold for purely transverse matrix conduction. The reference operator also retains longitudinal conduction in the matrix. For an anisotropic matrix, its strength at the front scale is $\chi\sqrt{\alpha_m}$; under isotropy it reduces to $\chi$ (Eq. (2.39)), independently of $L$ at fixed $u_0$. The no-flux longitudinal boundaries cancel the matrix contribution to the zeroth thermal balance, but its first spatial moment retains the boundary term $\kappa_{Lm}[\Theta_m(0,z,t) - \Theta_m(L,z,t)]$. This cancellation alone therefore does not extend the closed front moments to the full finite-domain model. The local field depends on $\kappa_{Lm}$ through the matrix symbol $\gamma = \sqrt{(s+\kappa_{Lm}q_x^2)/\kappa_{Tm}}$ of Eq. (2.25). At the semi-infinite exchange-front scale, its leading correction is $O(\alpha_m\chi^2)$, an asymptotic estimate rather than a uniform finite-domain bound. The profile, BTC and moment formulas are used for $\chi \ll 1$; the reference transport configurations have $\chi \approx 2\times10^{-2}$.

The hydraulic memory functional of Eq. (3.5) is the outlet-censored counterpart of the mean,

$$\hat{U}_1(\hat{s}) = -\frac{1}{\hat{s}}\,\frac{1-\exp\left(-\left(\hat{s}+\mathcal{A}\sqrt{\hat{s}}\right)\right)}{\hat{s}+\mathcal{A}\sqrt{\hat{s}}}, \tag{A.5}$$

so that $-U_1$ is the mean with its downstream contribution removed. $-U_1$ approaches unity algebraically, $1+U_1 \sim \mathcal{A}/(2\sqrt{\pi\tau})$ as $\tau\to\infty$; $U_1(0)=0$. The higher perturbative orders require no additional partial differential equation solves in this limit: the space–time impulse propagator $G_L$ associated with Eq. (A.2), that is $\partial_\tau S_L^{(0)}$, is the Lévy–Smirnov density carried by the advective delay,

$$G_L(\xi,\tau) = \frac{\mathcal{A}\xi}{2\sqrt{\pi}\,(\tau-\xi)^{3/2}}\exp\left[-\frac{\mathcal{A}^2\xi^2}{4(\tau-\xi)}\right]\mathrm{H}(\tau-\xi). \tag{A.6}$$

The first-order fields, the second-order functional $U_2$ of Eq. (3.10), and the first-order moments follow from Duhamel convolutions for the fields and spatial quadratures for the hydraulic averages in $U_2$.

The finite matrix breaks this structure in one specific place: the half-order exchange symbol in Eq. (A.1), $\mathcal{A}\sqrt{\hat{s}}$, is replaced by the capacity-limited family of Section 2.4, for which the Laplace inversion is no longer elementary.

## Appendix B. Finite-matrix similarity and Lauwerier crossover

This appendix derives the similarity laws used in Section 5.7 and documents the numerical controls behind the comparison shown there. The isoviscous unit-step problem is posed on an untruncated downstream domain, with rapid transverse equilibration ($Gz \ll 1$), negligible longitudinal diffusion in fluid and matrix, and strong matrix uptake ($C_m \gg 1$); the last condition yields the explicit similarity laws. Fig. 10 shows these laws; independent Laplace inversions check the moment laws in panels (e,f) within the same limit model (Appendix D.3). Use the dimensionless Laplace variable and closure indices introduced in Sections 4.3.2 and 3.1, with $S_j$ and $\Psi_j$ the front function and response symbol for closure $j$. The transformed step response can be written

$$\hat{S}_j(\xi,\hat{s}) = \frac{1}{\hat{s}}\exp[-\xi\Psi_j(\hat{s})], \tag{B.1}$$

where the dimensionless matrix exchange symbols are

$$\mathcal{K}_e(\hat{s}) = \mathcal{A}\sqrt{\hat{s}}\tanh\left(\pi\sqrt{\Pi_1\hat{s}}\right), \tag{B.2}$$

$$\mathcal{K}_t(\hat{s}) = \mathcal{A}\,\frac{\hat{s}}{\sqrt{\hat{s}+(\pi^2\Pi_1)^{-1}}}, \tag{B.3}$$

with $\mathcal{A}$ from Eq. (2.35). Both symbols reduce to $\mathcal{A}\sqrt{\hat{s}}$ at high frequency and to $C_m\hat{s}$ at low frequency.

### B.1. Full-field and breakthrough-curve similarity

Set $s_r = \Pi_1\hat{s} = st_1$, the Laplace variable conjugate to $r$ of Section 5.7. The scaled step-response field is given by Eq. (5.4). Across the crossover, $\mathcal{K}_j/\hat{s} = O(C_m)$, so $\Psi_j \sim \mathcal{K}_j$ as $C_m\to\infty$ at fixed $s_r$. The Laplace transform with respect to $r$ becomes, with scaled exchange symbols $\mathfrak{K}_j$,

$$\hat{\mathcal{T}}_j(\eta,s_r) = \frac{1}{s_r}\exp[-\eta\mathfrak{K}_j(s_r)],$$
$$\mathfrak{K}_e(s_r) = \sqrt{s_r}\tanh(\pi\sqrt{s_r}), \qquad \mathfrak{K}_t(s_r) = \frac{s_r}{\sqrt{s_r+\pi^{-2}}}. \tag{B.4}$$

Eq. (B.4) is a two-variable similarity law for the full step response.

At a detector $\xi_d$ the BTC is the cut

$$\mathcal{B}_j(r;\eta_d) = \mathcal{T}_j(\eta_d,r), \qquad \hat{\mathcal{B}}_j(s_r;\eta_d) = \frac{1}{s_r}\exp[-\eta_d\mathfrak{K}_j(s_r)], \qquad \eta_d = \frac{\mathcal{A}\xi_d}{\sqrt{\Pi_1}}. \tag{B.5}$$

BTCs therefore collapse against $r$ only when the scaled detector coordinate $\eta_d$ is held fixed. At the physical outlet $\xi_d = 1$, changing $\Pi_1$ also changes $\eta_d = \mathcal{A}/\sqrt{\Pi_1}$; the resulting curves are different cuts of one master surface, rather than a one-variable collapse. For example, with $\mathcal{A} = 14.18$, $\eta_d = 1.42$, 0.45 and 0.14 at $\Pi_1 = 10^2$, $10^3$ and $10^4$; the master field itself does not depend on $\mathcal{A}$.

At early time, $\mathfrak{K}_j(s_r) \sim \sqrt{s_r}$ and Eq. (B.4) gives the Lauwerier master response

$$\mathcal{T}_L(\eta,r) = \mathrm{erfc}\left(\frac{\eta}{2\sqrt{r}}\right). \tag{B.6}$$

At late time,

$$\mathfrak{K}_e(s_r) = \pi s_r - \frac{\pi^3}{3}s_r^2 + O(s_r^3), \qquad \mathfrak{K}_t(s_r) = \pi s_r - \frac{\pi^3}{2}s_r^2 + O(s_r^3), \tag{B.7}$$

so both have the same leading capacitive delay, $\mathcal{T}_j \sim \mathrm{H}(r-\pi\eta)$, while their first corrections differ.

### B.2. Mean and variance similarity

For the front on the half-line, integration of Eq. (B.1) gives the raw-moment transforms

$$\widehat{\mathcal{M}}_j(\hat{s}) = \frac{1}{\hat{s}\Psi_j(\hat{s})}, \qquad \hat{m}_{2,j}(\hat{s}) = \frac{2}{\hat{s}\Psi_j^2(\hat{s})}. \tag{B.8}$$

Using the same strong-uptake limit and scaled time $r$, introduce

$$\mathcal{M}_j(\tau) = \frac{\sqrt{\Pi_1}}{\mathcal{A}}\mathcal{F}_j(r). \tag{B.9}$$

The transforms of $\mathcal{F}_j$ are

$$\hat{\mathcal{F}}_e(s_r) = \frac{\coth(\pi\sqrt{s_r})}{s_r^{3/2}}, \qquad \hat{\mathcal{F}}_t(s_r) = \frac{\sqrt{s_r + \pi^{-2}}}{s_r^2}. \tag{B.10}$$

The exact-slab transform can be inverted by expanding the hyperbolic cotangent into images,

$$\mathcal{F}_e(r) = 2\sqrt{\frac{r}{\pi}} + 4\sum_{\ell=1}^{\infty}\left[\sqrt{\frac{r}{\pi}}\exp\left(-\frac{\ell^2\pi^2}{r}\right) - \ell\pi\,\mathrm{erfc}\left(\frac{\ell\pi}{\sqrt{r}}\right)\right], \tag{B.11}$$

whereas the tempered transform has the elementary inversion

$$\mathcal{F}_t(r) = \pi\left(\frac{1}{2} + \frac{r}{\pi^2}\right)\mathrm{erf}\left(\frac{\sqrt{r}}{\pi}\right) + \sqrt{\frac{r}{\pi}}\exp\left(-\frac{r}{\pi^2}\right). \tag{B.12}$$

Here $\mathrm{erf} = 1 - \mathrm{erfc}$ is the error function. Applying the same rescaling to the second relation in Eq. (B.8) gives

$$m_{2,j}(\tau) = \frac{\Pi_1}{\mathcal{A}^2}\mathcal{G}_j(r), \qquad \hat{\mathcal{G}}_e(s_r) = \frac{2\coth^2(\pi\sqrt{s_r})}{s_r^2}, \qquad \hat{\mathcal{G}}_t(s_r) = \frac{2(s_r + \pi^{-2})}{s_r^3}. \tag{B.13}$$

The tempered inversion is elementary,

$$\mathcal{G}_t(r) = 2r + \frac{r^2}{\pi^2}, \tag{B.14}$$

whereas the image expansion of $\coth^2(\pi\sqrt{s_r})$ yields

$$\mathcal{G}_e(r) = 2r + 8\sum_{\ell=1}^{\infty}\ell\left[(r + 2\ell^2\pi^2)\,\mathrm{erfc}\left(\frac{\ell\pi}{\sqrt{r}}\right) - 2\ell\sqrt{\pi r}\exp\left(-\frac{\ell^2\pi^2}{r}\right)\right]. \tag{B.15}$$

Consequently the central variance has the similarity form

$$\mathcal{V}_j(\tau) = m_{2,j}(\tau) - \mathcal{M}_j^2(\tau) = \frac{\Pi_1}{\mathcal{A}^2}\mathcal{H}_j(r), \qquad \mathcal{H}_j(r) = \mathcal{G}_j(r) - \mathcal{F}_j^2(r), \tag{B.16}$$

$$\mathcal{H}_L(r) = 2r - \left(2\sqrt{r/\pi}\right)^2 = 2r\left(1 - \frac{2}{\pi}\right), \tag{B.17}$$

where $\mathcal{H}_L$ is the scaled semi-infinite Lauwerier variance in the strong-uptake limit of Eqs. (A.3)–(A.4). Thus either $\mathcal{A}^2\mathcal{V}_j/\Pi_1$ or the relative departure

$$\delta_{\mathcal{V},j}(r) = \frac{\mathcal{H}_j(r)}{\mathcal{H}_L(r)} - 1 \tag{B.18}$$

collapses onto the same scaled time $r = \tau/\Pi_1$.

Expansion of the tempered law gives

$$\mathcal{F}_t(r) = 2\sqrt{\frac{r}{\pi}}\left[1 + \frac{r}{3\pi^2} + O(r^2)\right], \tag{B.19}$$

which immediately yields Eq. (5.3). Write $r_\delta = \tau_\delta/\Pi_1$ for the scaled threshold time. For a 2% relative departure, the asymptotic threshold is $r_{2\%} = 0.592$; direct solution of Eq. (B.12) gives 0.596. For reference, the corresponding 5% estimates are $r_{5\%} = 1.48$ and 1.50 from the expanded and complete laws, respectively.

The first image in Eq. (B.11) gives

$$\frac{\mathcal{F}_e(r) - 2\sqrt{r/\pi}}{2\sqrt{r/\pi}} \sim 2\exp\left(-\frac{\pi^2}{r}\right) - 2\pi\sqrt{\frac{\pi}{r}}\,\mathrm{erfc}\left(\frac{\pi}{\sqrt{r}}\right), \tag{B.20}$$

so the exact slab departs from Lauwerier only exponentially at first. Solving the complete image series gives $r_{2\%} = 3.83$ and $r_{5\%} = 5.25$, substantially later than the tempered thresholds. The variance behaves likewise: Eq. (B.18) gives

$$\delta_{\mathcal{V},t}(r) \sim \frac{3\pi - 8}{6\pi^3(1 - 2/\pi)}r \approx 0.0211r, \tag{B.21}$$

whereas the exact-slab variance departure is exponentially small.

As $r \to \infty$,

$$\mathcal{F}_e(r) = \frac{r}{\pi} + \frac{\pi}{3} + o(1), \qquad \mathcal{F}_t(r) = \frac{r}{\pi} + \frac{\pi}{2} + o(1), \tag{B.22}$$

and both means approach the common slope $\mathcal{M} \sim \tau/C_m$. This is the strong-uptake limit of the equilibrium-retarded slope $1/(1 + C_m)$.

The variance keeps a stronger memory of the closure. Expansion of Eqs. (B.14) and (B.15) gives

$$\mathcal{H}_t(r) = r - \frac{\pi^2}{4} + o(1), \qquad \mathcal{H}_e(r) = \frac{2r}{3} + \frac{\pi^2}{45} + o(1). \tag{B.23}$$

Accordingly, with $E_\mathcal{V} = \mathcal{V}_t/\mathcal{V}_e - 1$,

$$\delta_{\mathcal{V},t} \longrightarrow 0.37597, \qquad \delta_{\mathcal{V},e} \longrightarrow -0.082687, \qquad E_\mathcal{V} \longrightarrow \frac{1}{2}. \tag{B.24}$$

The $r$ collapse therefore extends to spreading, but the two closures do not become asymptotically equivalent.

The signed field difference (Fig. 10(b)) is positive near the advancing front; a negative wake appears later, from $r \approx 8$ at $\eta = 0.25$ to $r \approx 16$ at $\eta = 3$.

Both closures retain the same limiting responses, shown here for the tempered form:

$$\hat{H}_m^{\mathrm{temp}}(s) \sim \frac{1}{\sigma\sqrt{\kappa_{Tm}s}}, \qquad st_h \gg 1, \tag{B.25}$$

$$\hat{H}_m^{\mathrm{temp}}(s) \sim \frac{1}{\sigma h_m s}, \qquad st_h \ll 1. \tag{B.26}$$

The exact compliance has a storage pole at zero and discrete relaxation rates $\ell^2\pi^2/t_h$, with its first nonzero pole at $-\pi^2/t_h$. The tempered form has a branch point at $-1/t_h$ and a continuous spectrum above $t_h^{-1}$, represented numerically by positive exponentials. Both belong to the multirate mass-transfer family (Haggerty and Gorelick, 1995; Carrera et al., 1998; Babey et al., 2015); exponentially truncated memories are used in non-Fickian transport (Meerschaert et al., 2008; Tecklenburg et al., 2016; Dentz and Berkowitz, 2003). The corresponding step-temperature exchange kernel $k_e$, and the BTC symbol, instead have their first nonzero singularity at $-\pi^2/(4t_h)$; their leading rate ratio relative to tempering is $\pi^2/4$. These spectral rates do not determine a finite-time crossover threshold by themselves. The exact-to-tempered onset-time ratio is 6.4 at a 2% relative departure of the mean from Lauwerier, and 3.5 at 5%.

*Closure metrics and memory spectra.* The departures defined by Eq. (4.17) reach about 20% in time ($\varepsilon_H$), confined to the crossover band; $\varepsilon_D$ peaks at 8.81% on the positive Laplace axis. Both vanish toward their limiting responses. Their relation to observable errors is discussed in Section 4.4.

The closure difference in Fig. 9(c) follows from the dimensionless memory kernel $k(\tau)$ of Section 4.3.2, the matrix flux after a unit step in wall temperature. The semi-infinite memory $k_L \propto \tau^{-1/2}$ corresponds to the heavy-tailed, Lévy–Smirnov distribution of matrix residence times that underlies fractal and multirate mobile–immobile transport (Haggerty and Gorelick, 1995; Carrera et al., 1998; Schumer et al., 2003). The capacity-preserving closures of Section 3.1 truncate this tail differently. The tempered closure applies exponential truncation to the half-order memory, $k_t/k_L = \exp\left(-\tau/(\pi^2\Pi_1)\right)$: it modifies the memory at all times and decays on the scale $\pi^2 t_1$. The exact slab truncates by reflection at the no-flux boundary, $k_e/k_L = 1 + 2\sum_{\ell\geq 1}(-1)^\ell\exp\left(-\ell^2\pi^2\Pi_1/\tau\right)$: it is indistinguishable from the semi-infinite memory until heat reaches the boundary and then decays as $\exp\left(-\tau/(4\Pi_1)\right)$, the slowest slab exchange-memory mode. To hold the same capacity, the smooth exponential cut must extend to later times than the sharp reflective one; it places 56% of the capacity at relaxation times longer than $4t_1$, and so raises the mean residence time from $h_m^2/(3\kappa_{Tm})$ to $h_m^2/(2\kappa_{Tm})$ and the late variance by the same factor. Table B.1 gives the resulting signed departures in outlet timing and heat uptake. At the reference amplitude and $\lambda = 0$, tempering advances the 10% and half-breakthrough times by up to 14% (near $\Pi_1 \approx 3$) and 19% (near $\Pi_1 \approx 50$), delays the 99% time by up to 32%, and lowers heat uptake within $2t_r$ by up to 9% relative to the exact slab. At $\Pi_1 = 10^2$, tempered half-breakthrough precedes the semi-infinite value by 15%, whereas the exact slab lies within 1% (Section 4.2).

**Table B.1**
Signed departure (%) of the tempered closure from the exact slab at $\lambda = 0$, in the limit model with $\mathcal{A} = 14.18$. The times $t_{10}$, $t_{50}$ and $t_{99}$ correspond to $\Theta_{\text{out}} = 0.10$, 0.50 and 0.99. Heat uptake normalized by the inlet heat rate $\Gamma_f|\Delta T|bu_0$ is $Q(t) = \int_0^t [1 - \Theta_{\text{out}}(t')]\,\mathrm{d}t'$, evaluated at $t_r$ and $2t_r$ of Eq. (2.34). As $t \to \infty$, both closures take up the same total heat.

| $\Pi_1$ | $t_{10}$ | $t_{50}$ | $t_{99}$ | $Q(t_r)$ | $Q(2t_r)$ |
|---|---|---|---|---|---|
| 1 | −11.9 | −3.8 | +17.2 | −3.6 | −0.9 |
| 30 | −4.5 | −17.6 | +29.5 | −6.7 | −7.0 |
| $10^2$ | −1.4 | −14.8 | +31.9 | −6.3 | −8.7 |
| $10^4$ | 0.0 | −0.2 | +6.5 | −1.2 | −2.6 |

## Appendix C. Front moments: definition and moment representation

This appendix derives the moment representation of Eq. (4.12) and the error bound of Eq. (4.13) for the half-line limit model of Section 4.3.2.

*Setting.* The front function satisfies

$$\partial_\tau S + U(\tau)\,\partial_\xi S + \int_0^\tau k(\tau-\tau')\,\partial_{\tau'} S(\xi,\tau')\,\mathrm{d}\tau' = 0, \qquad \xi > 0,\ \tau > 0,$$
$$S(0,\tau) = 1, \qquad S(\xi,0) = 0. \tag{C.1}$$

Here $k$ is the matrix-memory kernel defined in Section 4.3.2, $*$ denotes causal convolution in $\tau$, and $U$ is continuous with $0 < U_- \le U(\tau) \le U_+ < \infty$. All three closures express $k$ as a positive sum or integral of decaying exponentials: for the slab, $k_e = 2C_m(\pi^2\Pi_1)^{-1}\sum_{\ell\ge1}\exp\big(-(\ell-1/2)^2\tau/\Pi_1\big)$, while $k_L \propto \tau^{-1/2}$ and $k_t = k_L\,\exp\big(-\tau/(\pi^2\Pi_1)\big)$ are exponential mixtures over rates $\omega > 0$, since $\tau^{-1/2} = \pi^{-1/2}\int_0^\infty \omega^{-1/2}\exp(-\omega\tau)\,\mathrm{d}\omega$; each kernel is integrable at $\tau = 0$. Finite-mode approximations therefore couple $S$ and the matrix temperatures through nonnegative coefficients.

*Propagator.* Since $\Psi(\hat{s})/\hat{s} = 1 + \hat{k}(\hat{s})$ is a Stieltjes function, $\Psi$ is a complete Bernstein function and $1/\Psi$ is Stieltjes (Schilling et al., 2012, Thms. 6.2 and 7.3). Thus $g$ is completely monotone: positive and nonincreasing. The limit $1/\Psi \to 0$ as $\hat{s} \to \infty$ excludes a Dirac contribution. The endpoint values of $g$ are $g(0^+) = 1$ and $g(\infty) = (1 + C_m)^{-1}$ for the finite-matrix closures, or zero for the semi-infinite closure.

*Front distribution.* The finite-mode system preserves $0 \le S \le 1$. For any shift $h > 0$, comparing $S$ with $S(\xi+h,\tau)$, whose inlet value is at most one, gives $\partial_\xi S \le 0$. The front vanishes beyond $\xi = \int_0^\tau U(\tau')\,\mathrm{d}\tau'$. Positive finite-mode kernels converge to $k$ in $L^1$ on every finite time window. For each spatial Laplace parameter $\upsilon \ge 0$, the transform $S_\upsilon(\tau) = \int_0^\infty \exp(-\upsilon\xi)\,S(\xi,\tau)\,\mathrm{d}\xi$ obeys the scalar Volterra equation $\partial_\tau S_\upsilon + k*\partial_\tau S_\upsilon = U(1 - \upsilon S_\upsilon)$, whose solution depends continuously on $k$ in $L^1$. Convergence of these transforms gives weak convergence of the front measures, preserving these properties. Hence $S(\cdot,\tau)$ is the survival function of a probability measure on the advective support. The moments of Eqs. (4.9)–(4.10) are its mean and variance, so $\mathcal{V} \ge 0$.

*Exact representation.* Integrating Eq. (C.1) over $\xi$, first directly and then after multiplication by $\xi$, gives in Laplace space $\Psi\widehat{\mathcal{M}} = \widehat{U}$ and $\Psi\widehat{m}_2 = 2\widehat{U\mathcal{M}}$, with zero initial values. Inverting gives Eq. (4.12). Subtracting $\mathcal{M}^2 = 2\iint_{0<\tau'<\tau''<\tau} g(\tau-\tau')g(\tau-\tau'')U(\tau')U(\tau'')\,\mathrm{d}\tau'\,\mathrm{d}\tau''$ gives

$$\mathcal{V}(\tau) = \iint_{0<\tau'<\tau''<\tau} C_\tau(\tau',\tau'')\,U(\tau')\,U(\tau'')\,\mathrm{d}\tau'\,\mathrm{d}\tau'', \tag{C.2}$$

with the kernel

$$C_\tau(\tau',\tau'') = 2\,g(\tau-\tau'')\left[g(\tau''-\tau') - g(\tau-\tau')\right] \ \ge 0, \tag{C.3}$$

where $C_\tau \ge 0$ because $g$ is nonincreasing and $\tau''-\tau' \le \tau-\tau'$. The mean is a positive linear functional of the velocity history, and the variance a quadratic form in it with nonnegative kernel; the subtraction of the mean is contained in the bracket of $C_\tau$.

*Semi-analytical thermo-viscous correction.* The expansion of Eq. (3.2) induces

$$S(\xi,\tau) = S^{(0)}(\xi,\tau) + \lambda S^{(1)}(\xi,\tau) + O(\lambda^2), \qquad S^{(j)}(\xi,\tau) = \overline{\Theta}_f^{(j)}(L\xi, t_a\tau), \tag{C.4}$$

where $j = 0, 1$ and $\overline{\Theta}_f^{(j)}$ denotes the aperture-averaged coefficient continued on the half-line within the limit model of Section 4.3.2. Applying Eq. (4.9) to each order gives $m_q = m_q^{(0)} + \lambda m_q^{(1)} + O(\lambda^2)$ and

$$\mathcal{V}^{(0)} = m_2^{(0)} - [\mathcal{M}^{(0)}]^2, \qquad \mathcal{V}^{(1)} = m_2^{(1)} - 2\mathcal{M}^{(0)}\mathcal{M}^{(1)}, \tag{C.5}$$

with the dimensional moments following from Eq. (4.11). Equivalently, by Eq. (4.12) with $U = 1 + \lambda U_1$, $\mathcal{M}^{(1)} = g*U_1$ and $m_2^{(1)} = 2g*(U_1\mathcal{M}^{(0)} + \mathcal{M}^{(1)})$, so both coefficients follow from the first-order hydraulic history without another transport solve.

*Consequences.* (i) For a stationary velocity $U \equiv U_\star$, $\mathcal{M} = U_\star\mathcal{M}^{(0)}$ and $\mathcal{V} = U_\star^2\mathcal{V}^{(0)}$ for every closure: a velocity change rescales the whole front. (ii) The positive-kernel representation of the first-order variance gives $\mathcal{V}^{(1)} = \iint_{0<\tau'<\tau''<\tau} C_\tau\,[U_1(\tau') + U_1(\tau'')]\,\mathrm{d}\tau'\,\mathrm{d}\tau''$, so $\mathcal{V}^{(1)}/\mathcal{V}^{(0)}$ is twice an average of $U_1$ with positive weights. As $\tau \to \infty$, $C_\tau$ assigns a vanishing fraction of its mass to any fixed early interval, so the ratio tends to $2U_1(\infty) = -2$, and $\mathcal{V}^{(0)} + \lambda\mathcal{V}^{(1)} < 0$ at late times for $\lambda > 1/2$. (iii) For two velocity histories $U$ and $U'$ driving the same transport problem, let primes identify moments driven by $U'$. Positivity of $g$ and $C_\tau$ gives, for $\tau > 0$,

$$\inf_{0\le\tau'\le\tau}\frac{U'(\tau')}{U(\tau')} \le \frac{\mathcal{M}'(\tau)}{\mathcal{M}(\tau)} \le \sup_{0\le\tau'\le\tau}\frac{U'(\tau')}{U(\tau')},$$
$$\left(\inf_{0\le\tau'\le\tau}\frac{U'(\tau')}{U(\tau')}\right)^2 \le \frac{\mathcal{V}'(\tau)}{\mathcal{V}(\tau)} \le \left(\sup_{0\le\tau'\le\tau}\frac{U'(\tau')}{U(\tau')}\right)^2. \tag{C.6}$$

The temperature depends on $\lambda$ only through the velocity, so the nonlinear solution and the reduced model of Eq. (3.11) are two such histories, which gives Eq. (4.13). The bound holds at every contrast. On a fixed finite time window, $\varepsilon_U = O(\lambda^3)$ when the hydraulic Taylor remainder is uniform on that window. It is evaluated a posteriori from the hydraulic histories.

## Appendix D. Numerical implementation and verification

This appendix collects the numerical methods, settings and checks used in the reference solutions and reduced-model comparisons. Literature contrast inputs are listed in Appendix D.5.

### D.1. Modal system and interface constraint

The fracture temperature is expanded in transverse cosine modes,

$$\Theta_f(x,z,t) = \sum_{n=0}^{N_z}\theta_n(x,t)\cos\left(\frac{n\pi z}{b}\right), \tag{D.1}$$

where $\theta_n$ are the transverse mode amplitudes and $N_z$ is the highest retained mode. Its projection onto Eq. (3.4) at leading order gives

$$A_0(x,t) = \theta_0^{(0)}(x,t) + \sum_{n=1}^{N_z}\frac{6(-1)^n}{n^2\pi^2}\theta_n^{(0)}(x,t). \tag{D.2}$$

Here $\theta_n^{(0)}$ are the leading-order coefficients of the same cosine expansion. The full projection is retained in all transport runs; numerical truncation is specified in Appendix D.3.

After projecting Eq. (2.2) onto these modes and discretizing in $x$, let $\boldsymbol{\theta}_n$ collect the grid values of mode $n$ and $\boldsymbol{\theta}$ stack these vectors. The modal equations take the form

$$\dot{\boldsymbol{\theta}} = B(u)\boldsymbol{\theta} + \boldsymbol{b}_f(u) + E_\theta \boldsymbol{q}_m, \tag{D.3}$$

where $B(u)$ contains longitudinal advection and diffusion together with the transverse decay rates $\kappa_{Tf}(n\pi/b)^2$, and $\boldsymbol{b}_f(u)$ is the inlet forcing. The vector $\boldsymbol{q}_m$ is the normalized signed interfacial heat flux into one adjacent matrix half-block, and $E_\theta$ distributes it over the transverse modes.

The state formulation uses closure-dependent positive weights $a_\ell$ and rates $\nu_\ell$, with $\ell = 0, \ldots, K-1$ and $K$ the number of retained matrix-memory states. For either finite closure, $\nu_0 = 0$. For the exact slab, Eq. (2.26) at $q_x = 0$ gives $\nu_\ell = \nu_\ell^{\mathrm{ex}}$ for $\ell \geq 1$, $a_0 = (\sigma h_m)^{-1}$, and $a_\ell = 2(\sigma h_m)^{-1}$ for $\ell \geq 1$. For tempering, let $\widetilde{a}_\ell$ and $\widetilde{\nu}_\ell$ be the positive quadrature weights and rates of $1/(\sigma\sqrt{\pi\kappa_{Tm}t})$. For $\ell = 1, \ldots, K-1$, the shifted rates and split weights are $\nu_\ell = \widetilde{\nu}_\ell + t_h^{-1}$, $a_\ell = \widetilde{a}_\ell \widetilde{\nu}_\ell/\nu_\ell$ and $a_0 = \sum_{\ell=1}^{K-1} \widetilde{a}_\ell t_h^{-1}/\nu_\ell$. Slab truncation preserves the capacitive weight exactly; the tempered weight is evaluated by quadrature. The semi-infinite reference uses the unshifted quadrature weights and rates, without a capacitive state. The corresponding memory states satisfy

$$\dot{\boldsymbol{\psi}}_\ell = \kappa_{Lm} D_{xx} \boldsymbol{\psi}_\ell - \nu_\ell \boldsymbol{\psi}_\ell + \boldsymbol{q}_m, \qquad \ell = 0, \ldots, K-1, \tag{D.4}$$

where $D_{xx}$ is the discrete longitudinal Laplacian; $\boldsymbol{\psi}$ stacks the matrix-memory states $\boldsymbol{\psi}_\ell$. The matrix-wall temperature is reconstructed as

$$\boldsymbol{\Theta}_{m,w} = \sum_{\ell=0}^{K-1} a_\ell \boldsymbol{\psi}_\ell. \tag{D.5}$$

Temperature continuity at the fracture–matrix interface gives the linear constraint

$$\sum_{n=0}^{N_z} (-1)^n \boldsymbol{\theta}_n = \sum_{\ell=0}^{K-1} a_\ell \boldsymbol{\psi}_\ell. \tag{D.6}$$

Differentiating Eq. (D.6) and substituting Eqs. (D.3) and (D.4) yields the explicit interfacial flux. Let $P$ be the matrix-state operator, block-diagonal with blocks $\kappa_{Lm} D_{xx} - \nu_\ell I$, where $I$ is the identity on the relevant grid or full-state space, $\otimes$ denotes the Kronecker product, and superscript $\mathsf{T}$ denotes transpose. With

$$\boldsymbol{c} = \left(1, -1, \ldots, (-1)^{N_z}\right)^{\mathsf{T}}, \qquad \boldsymbol{a} = (a_0, \ldots, a_{K-1})^{\mathsf{T}}, \tag{D.7}$$

the result is

$$\boldsymbol{q}_m = \frac{(\boldsymbol{c}^{\mathsf{T}} \otimes I)\left[B(u)\boldsymbol{\theta} + \boldsymbol{b}_f(u)\right] - (\boldsymbol{a}^{\mathsf{T}} \otimes I) P \boldsymbol{\psi}}{(1 + 2N_z)/b + \sum_{\ell=0}^{K-1} a_\ell}. \tag{D.8}$$

The factor $(1 + 2N_z)/b$ follows from the cosine projection of the wall flux: the zeroth fracture mode receives $-\boldsymbol{q}_m/b$, whereas each mode $n \geq 1$ receives $-2(-1)^n \boldsymbol{q}_m/b$; the factor is fixed by the differentiated interface constraint and is not a penalty parameter.

The reduced system preserves the wall-temperature constraint whenever the initial and history states satisfy Eq. (D.6).

Collect the unknowns in $\boldsymbol{Y} = (\boldsymbol{\theta}, \boldsymbol{\psi})^{\mathsf{T}}$ and eliminate $\boldsymbol{q}_m$ through Eq. (D.8) to obtain

$$\dot{\boldsymbol{Y}} = \boldsymbol{F}(\boldsymbol{Y}) = \mathcal{L}(u(\boldsymbol{Y}))\boldsymbol{Y} + \boldsymbol{f}(u(\boldsymbol{Y})), \tag{D.9}$$

with $\mathcal{L}$ the sparse semi-discrete transport operator and $\boldsymbol{f}$ the boundary forcing. Velocity enters only through advection, so $\mathcal{L}_u = \partial\mathcal{L}/\partial u$ and $\boldsymbol{f}_u = \partial \boldsymbol{f}/\partial u$ are independent of $u$; write $\mathcal{L}_0 = \mathcal{L}(u_0)$ and $\boldsymbol{f}_0 = \boldsymbol{f}(u_0)$. The exponential mobility of Eq. (2.22) gives $m_H > 0$ for bounded temperature and finite $\lambda$, so $u(\boldsymbol{Y})$ is smooth and the semi-discrete problem has a locally unique trajectory. The first-order expansion of Eq. (3.5) uses the leading-order state to determine velocity, so both orders share the constant operator $\mathcal{L}_0$ (Eqs. (D.12)–(D.14)).

Three reductions therefore act on distinct objects: the modal matrix replaces a two-dimensional field by $K$ exponential states, the elimination of the interface constraint gives ordinary differential equations, and the expansion makes hydraulic feedback explicit. The exponential representation and numerical tolerance determine the state count, as quantified in Appendix D.4 (Table D.4).

### D.2. Time integration and nonlinear solution

In the uncompressed slab expansion, the matrix eigenvalues $-\nu_\ell$ spread as $\ell^2$ and the stiffness ratio grows with the square of the spectral cutoff. This scaling does not describe compressed exponential representations.

Time integration uses backward differentiation formulas (BDF). The solver for hydraulic histories uses backward Euler (BDF1) with Newton iteration. Outlet maps and coupled viscosity-law comparisons use variable-step BDF2 with BDF1 startup. The memory-integration benchmark also uses BDF2. The moment evaluator instead applies Laplace inversion and quadrature to the velocity histories. At step $j$, $\boldsymbol{Y}_j$ is the discrete state and $\Delta t_j$ the time increment.

Let $\boldsymbol{g}_u = \nabla_{\boldsymbol{Y}} u$ be the full-state hydraulic gradient. With $w_i$ the longitudinal quadrature weights, $\sum_i w_i = L$, and $\theta_{n,i}$ the $n$th transverse mode at station $x_i$, Eqs. (2.13) and (2.15) give $\partial m_H/\partial\theta_{n,i} = -\lambda\langle\exp(-\lambda\Theta_f)\cos(n\pi z/b)\rangle_H$. Differentiating Eq. (2.19) gives

$$g_{u,n,i} = \frac{\partial u}{\partial \theta_{n,i}} = -\frac{\lambda u^2}{u_0 L}\frac{w_i}{m_H^2(x_i)}\left\langle \exp\left(-\lambda\Theta_f(x_i,z,t)\right)\cos\frac{n\pi z}{b}\right\rangle_H, \tag{D.10}$$

while $\partial u/\partial\boldsymbol{\psi} = 0$. The vector-field Jacobian is

$$\frac{\partial \boldsymbol{F}}{\partial \boldsymbol{Y}} = \mathcal{L}(u) + \left(\mathcal{L}_u \boldsymbol{Y} + \boldsymbol{f}_u\right)\boldsymbol{g}_u^{\mathsf{T}}, \tag{D.11}$$

that is, a rank-one hydraulic correction to the sparse transport operator.

With $\boldsymbol{Y} = \boldsymbol{Y}^{(0)} + \lambda\boldsymbol{Y}^{(1)} + O(\lambda^2)$ and $u = u_0(1 + \lambda U_1)$ in Eq. (D.9),

$$\dot{\boldsymbol{Y}}^{(0)} = \mathcal{L}_0 \boldsymbol{Y}^{(0)} + \boldsymbol{f}_0, \tag{D.12}$$

$$\dot{\boldsymbol{Y}}^{(1)} = \mathcal{L}_0 \boldsymbol{Y}^{(1)} + u_0 U_1(t)\left(\mathcal{L}_u \boldsymbol{Y}^{(0)} + \boldsymbol{f}_u\right), \tag{D.13}$$

with $U_1(t)$ from Eq. (3.5). Once $\boldsymbol{Y}^{(0)}$ is known, the forcing is assigned and the first-order Jacobian is

$$\left.\frac{\partial \dot{\boldsymbol{Y}}^{(1)}}{\partial \boldsymbol{Y}^{(1)}}\right|_{\boldsymbol{Y}^{(0)}} = \mathcal{L}_0, \tag{D.14}$$

in contrast with the full Jacobian of Eq. (D.11).

For the solver of hydraulic histories, the backward Euler residual is

$$\boldsymbol{R}_j(\boldsymbol{Y}) = \boldsymbol{Y} - \boldsymbol{Y}_{j-1} - \Delta t_j \boldsymbol{F}(\boldsymbol{Y}). \tag{D.15}$$

The Newton matrix is

$$\boldsymbol{J}_j = I - \Delta t_j \frac{\partial \boldsymbol{F}}{\partial \boldsymbol{Y}} = \left(I - \Delta t_j \mathcal{L}(u)\right) - \Delta t_j\left(\mathcal{L}_u \boldsymbol{Y} + \boldsymbol{f}_u\right)\boldsymbol{g}_u^{\mathsf{T}}. \tag{D.16}$$

The MATLAB driver uses inexact Newton iteration with the generalized minimal residual method (GMRES), preconditioned by an equilibrated LU factorization of the sparse part, recomputed at each Newton correction. With an exact factorization, GMRES requires at most two Krylov directions in exact arithmetic for a nonsingular system. Sherman–Morrison supplies the correction when GMRES fails or its relative unpreconditioned residual is nonfinite or exceeds $\max(\eta_{\mathrm{GMRES}}, 10^{-6})$, where $\eta_{\mathrm{GMRES}}$ is the current GMRES tolerance. Define the sparse part $\mathsf{P}_j = I - \Delta t_j\mathcal{L}(u)$ and the rank-one factor $\boldsymbol{v}_j = \Delta t_j(\mathcal{L}_u\boldsymbol{Y} + \boldsymbol{f}_u)$, so that $\boldsymbol{J}_j = \mathsf{P}_j - \boldsymbol{v}_j\boldsymbol{g}_u^{\mathsf{T}}$. With $\boldsymbol{z}_j = \mathsf{P}_j^{-1}(-\boldsymbol{R}_j)$ and $\boldsymbol{w}_j = \mathsf{P}_j^{-1}\boldsymbol{v}_j$, the Newton correction is $\delta\boldsymbol{Y} = \boldsymbol{z}_j + \boldsymbol{w}_j(\boldsymbol{g}_u^{\mathsf{T}}\boldsymbol{z}_j)/(1 - \boldsymbol{g}_u^{\mathsf{T}}\boldsymbol{w}_j)$; the code rejects a denominator smaller than $10^{-12}$ in magnitude.

The perturbative factorization depends only on the time-step coefficients, so it is shared by both orders and is independent of contrast. On the logarithmic grids it is rebuilt at each changed step; reuse across steps requires unchanged coefficients.

**Table D.1**
Numerical settings (top) and verification checks (bottom). There are $N_z+1$ fluid modes, including $n=0$; matrix states include the capacitive mode; $t_{\min}$ and $t_{\max}$ are the first and last output times, with $\tau_{\min} = t_{\min}/t_a$. Checks give the largest change under refinement or the largest difference from the stated reference.

| Quantity | Histories for front moments | Outlet validity maps |
|---|---|---|
| Cases | $\Pi_1 = 1,\ 10^2,\ 10^4$ | 65 values of $\lambda \in [-2, 2.4]$; 49 of $\Pi_1 \in [10^{-3}, 10^5]$; nonuniform, including $\lambda = \pm 0.6, \pm 1$ and $\Pi_1 = 1.038,\ 103.8,\ 10\,380$ |
| Fracture domain | $0 \le x \le L$ | $0 \le x \le 3L$ |
| Spatial mesh | 98 geometric cells; ratio 1.07; nominal first $10^{-4}L$, cap $0.05L$ | 192 cells; sinh stretching 3 |
| Fluid modes | $N_z = 1$ | $N_z = 3$ |
| Time grid | $\tau_j = 10^{-5+j/32} \le 3\times10^4$ | 481 log samples, $4.304\times10^{-4}$–$4.304\times10^3$ |
| Time stepping | Backward Euler (BDF1) | Variable-step BDF2; BDF1 startup |
| Residual tolerance | $10^{-6}$ | $2\times10^{-11}\max(1, \lvert U\rvert)$, hydraulic |
| Iteration limit | 150 | 16 Newton, then bracketed root |
| Safeguard | Line search to 1/65 536 | Root bracket; see text |
| Memory spacing | 0.28 | 0.3 |
| Matrix memory states | Set by the kernel interval | 70 |
| Kernel interval | $[t_{\min}/20, t_{\max}]$ | $[t_{\min}/20, t_{\max}]$ |

| Check | Result |
|---|---|
| Maps, $\Pi_1 = 1.038$: doubled cells; doubled samples | paired RMS $\le$1.9%; <0.3% |
| Map backend against archived MATLAB curves | $\le 1.5\times10^{-8}$ |
| History velocity, two finest levels | $\le$1.29% ($\Pi_1 = 1$), <0.74% ($10^2$, $10^4$) |
| Nonlinear moments, two finest levels, $\mathcal{M} \le 40$ | variance $\le$ 1.78% ($\Pi_1 = 1$), $\le$1.26%; mean <0.86% |
| Paired reconstruction errors | $\le$0.06 (successive), 0.09 (overall) percentage points |
| Moment quadrature against multiprecision | $10^{-12}$ (mean); $3\times10^{-5}$ (variance) |
| Extrapolated field moments, $\Pi_1 = 10^2$, $0.3 \le \mathcal{M} \le 5$ | <1.6% (mean); <4.3% (variance) |
| Similarity moments, Fig. 10(e,f) | Talbot order $< 10^{-20}$; laws $<10^{-8}$, $0.02 \le r \le 10^3$ |
| Vogel runs: doubled space; doubled time; tails | <0.6%; <0.04%; <3.3% |
| Table 5, tempered $E_s$: cells; samples; quadrature | <0.3%; <0.02%; <0.03% |
| Table 5, slab: quadrature; cells; time; $N_z$ | $<9\times10^{-6}$; <2.5%; <0.03%; <0.02% |
| Fixed-rate Vogel quadrature | $<2\times10^{-6}$ percentage points |
| Interface flux; energy partition | $<10^{-10}$; $3.4\times10^{-6}$ |
| Semi-infinite reference, $\Pi_1 = 10^4$, $\lambda = 0$ | outlet RMS $2.6\times10^{-3}$ |

### D.3. Numerical settings and verification

Table D.1 lists the settings of the two production drivers, which integrate the same modal equations, and the refinement and consistency checks behind the reported results.

The memory spacing $\Delta\zeta$ is the logarithmic quadrature step in the square-root rate, with nodes $\zeta_\ell$; the kernel interval is its construction window. The unshifted quadrature uses $\tilde{\nu}_\ell = \exp(2\zeta_\ell)$ and $\tilde{a}_\ell = 2\Delta\zeta\,\exp(\zeta_\ell)/(\pi\sigma\sqrt{\kappa_{Tm}})$, with uniformly spaced nodes from $\ln\sqrt{10^{-8}/t_{\max}}$ to $\ln\sqrt{36/(t_{\min}/20)}$ (last node not exceeding the upper bound). The spectral slab retains $N = \max[8, \lceil\sqrt{36\Pi_1/(\tau_{\min}/20)}\rceil]$ nonzero modes and the capacitive state. In the maps, quadrature changes the tempered capacity by at most 0.084% for $\Pi_1 \ge 1$ and by 0.50% at the smallest mapped thickness.

The history solver uses upwind advection and centered diffusion on an inlet-refined mesh and normalizes the Newton residual by the larger of one and the norm of the BDF history vector. Its three refinement levels take the inlet cell from $4\times10^{-4}L$ to $10^{-4}L$ (growth ratios 1.28, 1.14, 1.07; cell caps $0.2L$, $0.1L$, $0.05L$) and the output step from 8 to 32 per decade. Paired comparisons use identical grids, and front moments use the identities of Section 4.3.2. The map faces are $x_j = 3L\sinh(3j/192)/\sinh 3$. The map domain extends to $3L$ to keep its downstream condition away from the detector: the outlet temperature is sampled at $x = L$, the hydraulic resistance is integrated over $0 \le x \le L$, and a zero-gradient condition is applied at $x = 3L$. This differs from the outlet conditions of Section 2.1 only through longitudinal conduction across $x = L$, which is weak at the reference configuration ($u_0 L/\kappa_{Lf} \approx 7\times10^6$, $\chi = 0.02$). For $\Pi_1 < 1$, output intervals intersecting $[0.2(1+C_m)\exp(-2),\ 3(1+C_m)\exp(2.4)]$ are divided into up to eight logarithmic substeps to resolve the advective front. A Schur complement eliminates the memory states at each step, retaining both longitudinal-conduction operators; the hydraulic closure is solved by scalar Newton iteration with Brent's method as fallback, and the coupled residual is checked against $5\times10^{-7}$ at every step. The semi-infinite map reference is independent of $h_m$, so its 65 contrast histories serve every thickness. The semi-infinite outlet reference of the Roubinet et al. (2012) class uses 176 logarithmic samples over $10^{-4} \le \tau \le 10^3$ for the RMS check. Fig. 6 plots the retained 251-point reference through $\tau = 10^6$. The tail check of the viscosity-law runs applies where survival exceeds $10^{-4}$.

The hydraulic curves of Fig. 4 use 24 cosine-spaced cells, $N_z = 3$ and 201 logarithmic times over $10^{-3} \le \tau \le 10^5$; the breakthrough and energy curves of Figs. 6 and 7 use 320 uniform cells, $N_z = 5$ and 229 times over $10^{-4} \le \tau \le 10^8$. Both use BDF1, memory spacing 0.28 and tolerance $10^{-6}$, with 80 Newton iterations for the energy runs. At $\lambda = \pm0.6$, half of the velocity change takes 17–19 $t_a$ ($\Pi_1 = 1$) or 42–47 $t_a$ ($\Pi_1 = 10^2, 10^4$); the 90% time is 1.5–2.7 times longer under cooling than heating. Fig. 2(b,d) shows the extended viscosity-law histories over $1 \le \tau \le 10^4$; the tabulated metrics retain the 481-point window of Table D.1. Half-breakthrough crossings are interpolated linearly in temperature against $\log\tau$.

The discrete inlet flux uses the same advection–diffusion operator as the fracture solution. The verification checks refine each driver and each derived comparison, compare with independent references and verify conservation; Eq. (D.11) is also checked against directional finite differences. The checked spatial and temporal RMS changes at $\Pi_1 = 103.8$ and 10380 are smaller than at 1.038. Absolute moments and paired reconstruction errors are checked separately, so the stated reduced-model accuracy does not rely on grid cancellation alone.

**Table D.2**
Representative outlet discrepancies from the validity-map dataset, on the stated 481-point window. Classical: constant viscosity and semi-infinite matrix; first order: strict Taylor approximation with the same tempered matrix closure as the nonlinear reference. Numerical settings are given in Table D.1.

| $\Pi_1$ | $\lambda$ | Classical | | First order | |
|---|---|---|---|---|---|
| | | RMS | $E_\infty$ | RMS | $E_\infty$ |
| 1.038 | −0.6 | 0.2622 | 0.7376 | 0.0172 | 0.0789 |
| 1.038 | +0.6 | 0.1934 | 0.5773 | 0.0181 | 0.0858 |
| 103.8 | −0.6 | 0.0991 | 0.2124 | 0.0115 | 0.0332 |
| 103.8 | +0.6 | 0.0513 | 0.1290 | 0.0103 | 0.0267 |
| 10 380 | −0.6 | 0.0741 | 0.1819 | 0.0113 | 0.0327 |
| 10 380 | +0.6 | 0.0693 | 0.1592 | 0.0095 | 0.0261 |

*Outlet-error sampling.* For $E_s$ and $E_p$ of Section 5.2, $E_s$ = RMS $(\Theta^{\mathrm{NL}}_{\mathrm{out,finite}} - \Theta^{\mathrm{NL}}_{\mathrm{out,semi}})$ at matched $\lambda$, and $E_p = \mathrm{RMS}(\Theta^{\mathrm{NL}}_{\mathrm{out,finite}} - \Theta^{(1\mathrm{st})}_{\mathrm{out,finite}})$ with the same tempered closure. Here NL denotes the nonlinear reference, and finite and semi identify finite and semi-infinite matrix responses. For the compared normalized outlet temperatures, RMS = $[N_t^{-1}\sum_j(\Delta\Theta_{\mathrm{out},j})^2]^{1/2}$ and $E_\infty = \max_j|\Delta\Theta_{\mathrm{out},j}|$, with $N_t$ the number of output samples and $\Delta\Theta_{\mathrm{out},j}$ their difference at sample $j$. The RMS and maximum discrepancies use the 481-point logarithmic output grid of Table D.1. Both endpoints are included and every sample has equal weight; errors are normalized by $|T_{\mathrm{inj}} - T_0|$.

*Paired reconstruction errors.* At $\lambda = \pm 0.6$, the quadratic hydraulic form $1 + \lambda U_1 + \lambda^2 U_2$ gives $\varepsilon_U$ = 0.032–0.065, against 0.009–0.013 for the exponential form. The latter bounds moment errors by 0.9%–1.3% in the mean and 1.74%–2.65% in the variance via Eq. (4.13). At $\Pi_1 = 10^2$ and $|\lambda| = 1$, maximum mean and variance errors are 3.1% and 6.0% under cooling, and 4.3% and 8.3% under heating. At $|\lambda| = 1.5$, the respective pairs are 9.9% and 19.6%, and 14.9% and 27.3%. Refinement uses three levels at $\Pi_1 = 10^2$ and two at $\Pi_1 = 1$ and $10^4$. The sampled window is $1 \le \tau \le 2\times10^4$ with $\mathcal{M} \le 40$ for the nonlinear history, so its endpoint depends on thickness and contrast.

*Moment evaluation.* The identities of Eq. (4.12) are evaluated by product integration and fixed-Talbot inversion of $1/\Psi$ (Abate and Valkó, 2004). At the reference configuration, longitudinal fluid conduction is measured by $u_0 L/\kappa_{Lf} \approx 7\times10^6$ and matrix conduction by $\chi = 0.02$ (Eq. (2.39)); its leading variance correction in the semi-infinite exchange regime is quadratic in $\chi$ (Appendix A). The field-moment check of Table D.1, at $\lambda \in \{0, \pm 0.6\}$, uses first-order Richardson extrapolation on a $12.5L$ domain with downstream temperature below $10^{-2}$. It covers $0.3 \le \mathcal{M} \le 5$; discrepancies are reported in Section 4.3.2.

*Exact-slab comparisons.* The exact-slab entries of Table 5 keep the production modal solver and both longitudinal-conduction terms, changing only the finite-matrix compliance. Six-point block quadrature (Appendix D.4) with the kernel interval of Table D.1 uses 34, 40 and 52 states at the three thicknesses, with compliance error below $6.1\times10^{-5}$ relative. The transverse-mode check increases $N_z$ from 3 to 5 at $\Pi_1 = 103.8$ (Table D.1). Values of $E_s$ below $10^{-4}$ are reported only as upper bounds, since residual compliance and discretization differences can dominate that signal.

*Closure dependence of thermo-viscous shifts.* The heating and cooling runs were repeated with the compressed exact slab at $\Pi_1$ = 1, $10^2$, $10^4$ in both production drivers, using the map settings above and 40, 52 and 58 states in the history driver, whose kernel window is longer. Each closure uses its own nonlinear velocity history and propagator. Heating advances and broadens the front and cooling retards and sharpens it with both closures. At matched time, the relative changes $\mathcal{M}/\mathcal{M}^{(0)} - 1$ and $\mathcal{V}/\mathcal{V}^{(0)} - 1$ at $\lambda = \pm 0.6$ differ by at most 1.1 and 1.4 percentage points. Relative half-breakthrough shifts for $\lambda \in \{-0.6, -0.4, 0.6, 0.8\}$ differ by at most 2.9 percentage points at $\Pi_1 = 1$ and 1.4 at $\Pi_1 = 10^4$. At $\Pi_1 = 10^2$ the slab gives larger shifts, −53.6% at $\lambda = -0.6$ and +103.8% at $\lambda = 0.6$, against −49.6% and +93.4% with tempering, and +149.0% against +138.2% at $\lambda = 0.8$, because the tempered response leaves the semi-infinite regime earlier. For the viscosity-law comparison at $\Pi_1 = 103.8$, the exact slab gives half-breakthrough shifts of 21.8% (200 → 40 °C) and 37.1% (reverse), against 25.1% and 36.2% with tempering.

*Velocity-profile check.* The section-mean closure is compared with an isoviscous solution that advects heat with the plane-Poiseuille profile $\frac{3}{2}u(1 - z^2/b^2)$. Both solutions resolve transverse diffusion on 40 finite volumes clustered toward the wall, exchange heat with the exact finite slab and omit longitudinal conduction, so their difference isolates the velocity profile. They are solved in the Laplace domain and inverted numerically; doubling the transverse resolution leaves the tabulated values unchanged. Table D.3 compares the flux-weighted outlet temperature $\Theta^F_{\mathrm{out}} = \int_0^b v\,\Theta_f\,\mathrm{d}z / \int_0^b v\,\mathrm{d}z$ at $x = L$, which a production well samples, with the closure. The faster central fluid carries more of the heat flux, so the flux-weighted temperature arrives earlier, whereas the section-mean temperature, which sets the exchange and the front position, changes much less. The flux-weighted shift grows approximately in proportion to $Gz$. Varying the aperture at fixed properties and transit time, so that $\mathcal{A} = \sigma/\sqrt{\beta_T Gz}$ decreases as $Gz$ increases, changes the half-breakthrough shifts by at most 0.5 percentage points for $Gz \le 0.3$.

At finite contrast, viscosity changes $Gz$ through $U(t)$ and also distorts the profile. A coupled finite-volume reference includes both effects. It computes the lubrication profile from the local exponential viscosity $\mu_0 e^{\lambda\Theta}$ in every cell, keeps the discharge of Eq. (2.19) at the imposed pressure drop, and obtains the transverse velocity from the streamfunction, so that continuity holds exactly in the discrete fluxes; the exact slab is resolved in each column. The same code with the section-mean velocity gives the paired closure. Both use first-order upwind advection, backward Euler steps and Picard iteration on the velocity field. At $\lambda = 0$ the coupled code reproduces Table D.3(a) for $Gz$ = 0.017, 0.05 and 0.13 to within 0.01 percentage points in $\Delta\tau^{\mathrm{p}}_{50}$ and about 2% in RMS and $E_\infty$. Refining space and time separately changes the paired differences by at most 0.01 percentage points in $\Delta\tau^{\mathrm{p}}_{50}$ and 3% in RMS and $E_\infty$; a uniform temperature is preserved exactly, and energy is conserved to $10^{-11}$. Table D.3(b) covers seven selected contrasts spanning the map range at the reference $Gz = 0.017$, with the extreme contrasts repeated at $\Pi_1 = 1$ and $10^4$. In these cases $Gz\,U(\tau_{50}) \le 0.08$, and the shifts are smaller than those of panel (a) at comparable $Gz$, where $Gz$ is fixed from the start; they document the tested cases.

*Material-property and buoyancy estimates.* Constant density and thermal properties are approximations for modest temperature changes: between 55 and 80 °C at 0.1 MPa, IAPWS-95 gives decreases of 1.4% in density and 1.1% in volumetric heat capacity (IAPWS, 2018). Between 40 and 200 °C at 10 MPa the density falls by 12.6% and the volumetric heat capacity by about 6%, while the thermal conductivity rises by about 5% (Huber et al., 2012); the constant-property model is then a viscosity-isolating approximation. Buoyancy in a fracture inclined by $\alpha$ above the horizontal is negligible when $|\Delta\rho| g_{\mathrm{grav}} L |\sin\alpha|/\Delta P \ll 1$, where $\Delta\rho$ is the density contrast and $g_{\mathrm{grav}} = 9.81\ \mathrm{m\,s^{-2}}$: at the reference forcing ($L$ = 10 m, $\Delta P$ = 11.3 kPa), the ratios are $0.12|\sin\alpha|$ for 55–80 °C and $1.1|\sin\alpha|$ for 40–200 °C. Rescaling pressure with the initial-state Vogel viscosity to retain $u_0$ (Section 4.1) gives $\Delta P \approx 3.96$ and 1.51 kPa for 80 → 55 and 200 → 40 °C, respectively; the ratios become $0.34|\sin\alpha|$ and $8.2|\sin\alpha|$. Negligible buoyancy therefore requires the applied pressure to be large compared with $|\Delta\rho| g_{\mathrm{grav}} L|\sin\alpha|$; horizontal fractures satisfy this condition.

**Table D.3**
Velocity-profile check at the reference $\mathcal{A} = 14.18$: differences between profile-resolved and section-mean advection of the flux-weighted outlet temperature $\Theta^F_{out}$ (RMS and maximum $E_\infty$, normalized by $|T_{inj} - T_0|$; relative half-breakthrough shift $\Delta\tau^p_{50} = 100(\tau^{prof}_{50}/\tau^{mean}_{50} - 1)$, negative when earlier, where $\tau^{mean}_{50}$ refers to the section-mean closure, in which $\Theta^F_{out} = \Theta_{out}$); the section-mean column applies the same definition to $\Theta_{out}$. (a) Isoviscous Laplace-domain solution on 161 logarithmic samples of the window of Table D.1; the section-mean column is for $\Pi_1 = 10^2$, and values at $\Pi_1 = 10^4$ differ from those at $10^2$ by at most 0.1 percentage points in $\Delta\tau^p_{50}$. (b) Coupled finite-volume solution on the 481-sample window, with the exponential viscosity law at $Gz = 0.017$; $Gz\,U(\tau_{50})$ uses the velocity history and half-breakthrough time of the section-mean closure; $\lambda = \pm1.583$ corresponds to 200 ↔ 40 °C with Vogel endpoints. At $\Pi_1 = 10^4$ and $\lambda = 2.4$, half-breakthrough lies beyond the window.

(a) Isoviscous

| $Gz$ | $\Pi_1 = 1$ | | | $\Pi_1 = 10^2$ | | | Section mean |
|---|---|---|---|---|---|---|---|
| | RMS | $E_\infty$ | $\Delta\tau^p_{50}$ (%) | RMS | $E_\infty$ | $\Delta\tau^p_{50}$ (%) | $\Delta\tau^p_{50}$ (%) |
| 0.017 | 0.0011 | 0.0052 | −0.11 | 0.0007 | 0.0024 | −0.23 | +0.05 |
| 0.05 | 0.0031 | 0.0131 | −0.35 | 0.0021 | 0.0066 | −0.70 | +0.14 |
| 0.13 | 0.0067 | 0.0260 | −1.01 | 0.0054 | 0.0156 | −1.92 | +0.30 |
| 0.3 | 0.0129 | 0.0414 | −2.70 | 0.0124 | 0.0310 | −4.91 | +0.41 |
| 1 | 0.0387 | 0.148 | −11.7 | 0.0420 | 0.148 | −22.2 | −1.13 |

(b) Coupled, viscosity-dependent profile

| $\Pi_1$ | $\lambda$ | $Gz\,U(\tau_{50})$ | RMS | $E_\infty$ | $\Delta\tau^p_{50}$ (%) | Section mean $\Delta\tau^p_{50}$ (%) |
|---|---|---|---|---|---|---|
| $10^2$ | −2 | 0.073 | 0.0016 | 0.0089 | −0.26 | +0.20 |
| $10^2$ | −1.583 | 0.054 | 0.0013 | 0.0061 | −0.17 | +0.29 |
| $10^2$ | −0.6 | 0.026 | 0.0009 | 0.0031 | −0.21 | +0.17 |
| $10^2$ | 0 | 0.017 | 0.0007 | 0.0023 | −0.23 | +0.05 |
| $10^2$ | 0.6 | 0.011 | 0.0006 | 0.0018 | −0.18 | −0.02 |
| $10^2$ | 1.583 | 0.005 | 0.0005 | 0.0013 | −0.11 | −0.04 |
| $10^2$ | 2.4 | 0.002 | 0.0004 | 0.0010 | −0.10 | −0.06 |
| 1 | −2 | 0.080 | 0.0019 | 0.0123 | −0.40 | −0.08 |
| 1 | 2.4 | 0.002 | 0.0011 | 0.0048 | −0.24 | −0.21 |
| $10^4$ | −2 | 0.073 | 0.0016 | 0.0084 | −0.25 | +0.20 |
| $10^4$ | 2.4 | – | 0.0004 | 0.0009 | – | – |

### D.4. Memory-integration benchmark

Table D.4 compares three memory representations at $\lambda = 0$ and $\mathcal{A} = 14.1845$ in the rapid-transverse-equilibrium limit without longitudinal conduction. Independent Laplace inversions use the exact-slab symbol for both slab representations and the tempered symbol for the tempered closure, with a common 1% tolerance on moments, normalized outlet temperature and the 99% breakthrough time; model-form bias is excluded. The reported error is the maximum of relative moment errors for $\tau \geq 10^{-2}$, absolute dimensionless moment errors earlier, absolute normalized-temperature error and relative 99% breakthrough-time error. References include 100 extra samples across breakthrough. The compressed slab replaces each disjoint block of integer modes $[\ell, 4\ell - 1]$ (clipped at $N$) by six-point Gaussian quadrature of the uniform discrete measure; smaller blocks are kept exactly. Quadrature nodes are squared and divided by $\Pi_1$ to give rates, their weights are doubled, and the capacitive state is retained. The tempered benchmark uses the same positive quadrature construction as above with $\Delta\zeta = 0.4$.

Compressed slab and tempering have comparable costs at matched numerical accuracy in this benchmark. The spectral cost reflects that representation rather than the slab closure, and the benchmark isolates memory integration. Model-form departures are assessed separately in Table B.1 and Section 5.6.

**Table D.4**
Memory-integration benchmark on an AMD Ryzen 9 7940HX with 31.2 GiB RAM, Windows 11 and Python 3.10.6, using one numerical thread. Times are medians of three isolated runs after warm-up, excluding startup, kernel construction and output; the full timing range is below 5% of the median. Memory is peak private process memory, including the runtime. All cases have a 60 s/16 GiB resource budget; grids and time windows are unchanged when censored. States include the capacitive mode. All use variable-step BDF2 with 1024 uniform cells on $0 \leq \xi \leq 1$ and 3073 logarithmic times over $10^{-5} \leq \tau \leq 3 \times 10^6$, with local Schur elimination.

| $\Pi_1$ | Representation | States | Time (s) | Peak (MiB) | Error (%) |
|---|---|---|---|---|---|
| 1 | Slab spectral | 8 487 | >60 | 314.1[a] | – |
| 1 | Slab compressed | 40 | 0.212 | 115.4 | 0.128 |
| 1 | Tempered | 66 | 0.292 | 115.5 | 0.114 |
| $10^2$ | Slab spectral | 84 854 | >60 | 2114.4[a] | – |
| $10^2$ | Slab compressed | 52 | 0.255 | 115.4 | 0.125 |
| $10^2$ | Tempered | 66 | 0.298 | 115.5 | 0.114 |
| $10^4$ | Slab spectral | 848 530 | –[b] | –[b] | – |
| $10^4$ | Slab compressed | 58 | 0.274 | 115.1 | 0.125 |
| $10^4$ | Tempered | 66 | 0.293 | 115.8 | 0.114 |

[a] Peak observed before time censoring; achieved accuracy is unavailable.
[b] Censored before allocation: the three state buffers require 19.46 GiB, exceeding the common budget.

### D.5. Literature inputs and parameter context

Let $d_n$ denote fracture-normal spacing and $\mathcal{N}$ a normal distribution. The lognormal curve in Fig. 8b of Massiot et al. (2017b) is reconstructed as $\ln(d_n/\mathrm{m}) \sim \mathcal{N}(0.00619, 1.35916^2)$. These parameters come from its digitized vector centerline on the published logarithmic axes, fitted over cumulative fractions 0.01–0.95 (maximum CDF reconstruction error $5 \times 10^{-4}$). The extracted curve and calibration are archived alongside the separately digitized empirical points; this is not a fit to the original observations. The projection takes $h_m = d_n/2$ with the reference $\kappa_m$ and $t_a$ of Table 3, giving $\Pi_1 \approx 510\,(h_m/\mathrm{m})^2$. Table D.5 provides temperature-contrast context using the Vogel relation of Section 2.3; no matrix thickness or $\Pi_1$ is inferred for these studies.

*Fixed-rate benchmark inputs.* The nominal $\mathcal{A}$ follows from aperture, geometry, rate and thermal properties through Eq. (2.35); in the field the aperture is the least constrained of these inputs. The thesis inputs (Okoroafor, 2021, Chapter 4, Section 4.2.1) give $\mathcal{A} = 90.9$; the article's rock properties ($\rho_r = 2670$ kg m$^{-3}$, $c_{p,r} = 1600$ J kg$^{-1}$ K$^{-1}$) give $\mathcal{A} = 116.5$ under the same conversion. The thesis-based inputs are fracture length and width $L = W = 1000$ m, aperture $2b = 5$ mm, mass flow rate $\dot{m} = 40$ kg s$^{-1}$, $\rho_r = 2600$ kg m$^{-3}$, $c_{p,r} = 1000$ J kg$^{-1}$ K$^{-1}$, $k_r = 2.8$ W m$^{-1}$ K$^{-1}$, $\rho_f = 1000$ kg m$^{-3}$ and $c_{p,f} = 4200$ J kg$^{-1}$ K$^{-1}$ (Okoroafor, 2021). The joint fit to the retained outlet digitization and its window are reported in Section 5.8. The fitted amplitude is 6.4% below the thesis-based nominal value and 27% below the article-based value. The fixed-rate Vogel check integrates its viscosity on the same calibrated leading-order temperature profile as Fig. 11(c), using 4000 longitudinal points and 250 logarithmic times over 0.02–5.2 years; no transport rerun or refit is used. For $200 \to 40$ and $200 \to 60$ °C, maximum viscosity departures from Vogel are 29.3% and 18.3%, respectively. Eq. (4.15) bounds the corresponding pressure reductions at the same temperature field by 22.6% and 15.5%.

**Table D.5**
Literature temperatures, Vogel-derived viscosity contrasts and estimated Graetz numbers. Within each grouped row the viscosity contrasts follow the order of the injection temperatures. The order-of-magnitude estimate $Gz = b^2/(\kappa_{Tf} t_a)$ uses $\kappa_{Tf} = 1.4 \times 10^{-7}$ $m^2 s^{-1}$, the reported or calibrated aperture $2b$ and a transit time $t_a$, estimated from the reported geometry and rate for the simulations (direct-line dipole time for Guo et al. (2016)) and taken from solute arrivals in the field tests. For Hawkins et al. (2017) the range spans the calibrated uniform and channeled apertures at the first tracer arrival; for de La Bernardie et al. (2018) it spans the modal and minimum solute transit times for the planar aperture of 3.3 mm, whereas their alternative cylindrical channel lies outside the planar geometry. For Hoffmann et al. (2022) the calibrated apertures and the reported arrival times belong to different campaigns, so no value is given.

| Source | $T_0 \to T_{inj}$ (°C) | $\lambda$ | $Gz$ |
|---|---|---|---|
| Okoroafor and Horne (2022) | $120 \to 20, 40, 60$ | 1.472, 1.041, 0.700 | $3.6 \times 10^{-4}$ |
| | $150 \to 20, 40, 60$ | 1.711, 1.280, 0.939 | |
| | $200 \to 20, 40, 60$ | 2.013, 1.583, 1.241 | |
| | $250 \to 20, 40, 60$ | 2.236, 1.806, 1.465 | |
| Fox et al. (2015)[a] | $200 \to 65$ | 1.167 | – |
| Guo et al. (2016) | $200 \to 50$ | 1.403 | $2.0 \times 10^{-5}$ |
| Hawkins et al. (2017) | $11.7 \to 74$ | −1.185 | $6.0 \times 10^{-4}$–0.069 |
| de La Bernardie et al. (2018) | $16.7 \to 56$ | −0.791 | $9.0 \times 10^{-3}$–0.019 |
| Hoffmann et al. (2022) | $30 \to 50, 10$ | −0.382, 0.489 | – |

[a] The single-fracture results of Fox et al. (2015) are dimensionless; the temperature pair is the dimensional case with which Okoroafor and Horne (2022) reproduce them; no $Gz$ is given because the geometry of that case differs from the published example.

## Data availability

The MATLAB and Python codes and datasets used in this study are archived on Zenodo under a CC BY 4.0 license (https://doi.org/10.5281/zenodo.23301912).